\documentclass[journal]{IEEEtran}
\usepackage{graphicx}
\usepackage{multirow}
\usepackage{booktabs}
\usepackage{wasysym}
\usepackage{amssymb}
\usepackage{multirow}
\usepackage{colortbl}
\usepackage{soul}
\usepackage{color}
\usepackage{makecell}
\usepackage[table]{xcolor} 
\usepackage{xcolor}

\usepackage{hyperref}
\hypersetup{
	colorlinks=true,
	linkcolor=blue,
	urlcolor=blue,
	citecolor=blue
}

\begin{document}
\title{Dynamic Cross-Modal Feature Interaction Network for Hyperspectral and LiDAR Data Classification}
\author{
        Junyan Lin,
        Feng Gao, \textit{Member, IEEE},
        Lin Qi,
        Junyu Dong, \textit{Member, IEEE}, \\
        Qian Du, \textit{Fellow, IEEE},
        Xinbo Gao, \textit{Fellow, IEEE}
\thanks{This work was supported in part by the National Key Research and Development Program of China under Grant 2022ZD0117202, 
in part by the Natural Science Foundation of Shandong Province under Grant ZR2024MF020,
and in part by the Natural Science Foundation of Qingdao under Grant 23-2-1-222-ZYYD-JCH. ( \textit{Corresponding author: Feng Gao} )

Junyan Lin, Feng Gao, and Junyu Dong are with the School of Computer Science and Technology, Ocean University of China, Qingdao 266100, China.

Qian Du is with the Department of Electrical and Computer Engineering, Mississippi State University, Starkville, MS 39762 USA.

Xinbo Gao is with the Chongqing Key Laboratory of Image Cognition, Chongqing University of Posts and Telecommunications, Chongqing 400065, China.}
}

\markboth{IEEE Transcations on Geoscience and Remote Sensing}
{Shell}

\maketitle

\begin{abstract}

Hyperspectral image (HSI) and LiDAR data joint classification is a challenging task. Existing multi-source remote sensing data classification methods often rely on human-designed frameworks for feature extraction, which heavily depend on expert knowledge. To address these limitations, we propose a novel Dynamic Cross-Modal Feature Interaction Network (DCMNet), the first framework leveraging a dynamic routing mechanism for HSI and LiDAR classification. Specifically, our approach introduces three feature interaction blocks: Bilinear Spatial Attention Block (BSAB), Bilinear Channel Attention Block (BCAB), and Integration Convolutional Block (ICB). These blocks are designed to effectively enhance spatial, spectral, and discriminative feature interactions. A multi-layer routing space with routing gates is designed to determine optimal computational paths, enabling data-dependent feature fusion. Additionally, bilinear attention mechanisms are employed to enhance feature interactions in spatial and channel representations. Extensive experiments on three public HSI and LiDAR datasets demonstrate the superiority of DCMNet over state-of-the-art methods. Our code will be available at \url{https://github.com/oucailab/DCMNet}.

\end{abstract}


\IEEEpeerreviewmaketitle

\section{Introduction}

\IEEEPARstart{W}ITH the development of spectroscopy and imaging technologies, satellite sensors can capture a large number of high-resolution images that can be applied to environmental monitoring \cite{chong24tgrs}, vegetation mapping \cite{jiao24veg}, coastal wetland mapping \cite{cui22grsl}, etc. Among these images, hyperspectral images (HSIs) contain hundreds of contiguous spectral bands, which facilitates fine discrimination between different ground objects on the Earth's surface. Therefore, HSI classification has been the most active part of the research in the remote sensing society \cite{yu24tgrs}.

The spectral information in HSIs alone is often insufficient to accurately classify ground objects due to their complex composition. Specifically, the spectral information of ground objects is often complex, and thus leads to the phenomenon of ``same spectrum corresponds to multiple ground coverings" \cite{zhou22grsl}. The light detection and ranging (LiDAR) data are obtained by actively emitting light at the ground coverings and can create high-resolution models of ground elevation \cite{huang2024sunshine}. Therefore, LiDAR data provides complementary information for HSI in ground covering classification. The combination of HSI and LiDAR data takes advantage of the complementary multisource information, and improves the accuracy of ground object classification \cite{gu17grsl} \cite{wang23tgrs}. This multisource fusion paradigm has been effectively adopted in other fields \cite{huang2024v2x} \cite{huang2024l4dr}.

\begin{figure}[]
\centering
\includegraphics [width=3.2in]{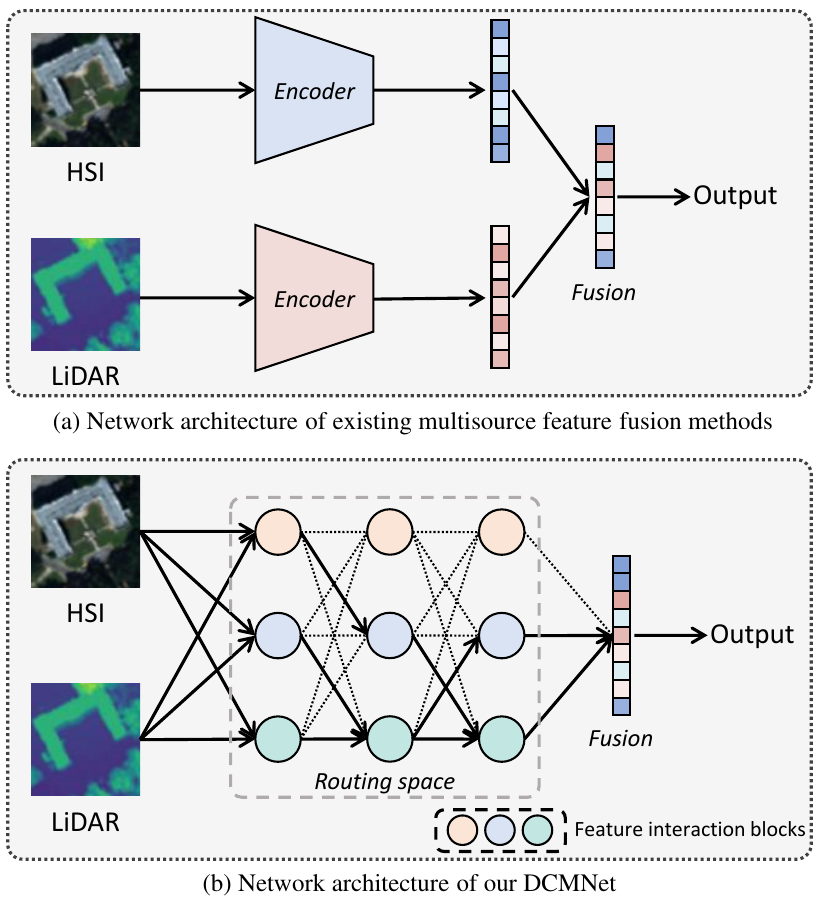}
\caption{Comparisons of the traditional method with the proposed DCMNet. (a) Existing multi-source feature fusion method. Their static framework lacks the adaptability to ground objects with semantic diversity. (b) The proposed DCMNet. Dynamic routing mechanism is introduced to learn data-dependent feature extraction paths.}
\label{fig_comp}
\end{figure}

Based on this concept, numerous studies have explored joint HSI and LiDAR classification, utilizing different fusion strategies. They can be categorized into two groups: decision-level fusion methods \cite{liao2014combining} \cite{bigdeli15} \cite{jia21decision} and feature-level fusion methods \cite{chen2017deep} \cite{xu2017multisource}  \cite{li2020a3clnn}. Decision-level fusion methods commonly use several classifiers individually. Afterwards, graph fusion \cite{xia18grsl} and active learning \cite{zhang2014ensemble} are used to integrate the classification results. In feature-level fusion methods, deep learning-based techniques have attracted wide attention recently. Self-guided attention \cite{dong22tgrs}, graph-enhanced network \cite{graph22tgrs}, cross-attention network \cite{gao2022tgrs}, and flexible-Mixup \cite{wang23tnnls} have achieved great success in learning and fusing multi-source features.

Although these pioneer works have made significant strides in addressing the challenges of HSI and LiDAR data joint classification, they still face several critical limitations, particularly when dealing with data heterogeneity, high dimensionality, and the integration of spatial and spectral information. Data heterogeneity arises from the fundamental differences between HSI and LiDAR data, as HSI captures detailed spectral information across numerous bands, while LiDAR provides precise spatial and elevation data. These distinct characteristics make it difficult for traditional methods to effectively fuse and exploit complementary information from both modalities. Additionally, the high dimensionality of HSI data presents computational challenges, especially when combined with the complexity of integrating spatial information from LiDAR. Current methods often struggle to fully utilize this wealth of information, leading to suboptimal classification results. Furthermore, existing methods frequently fail to properly integrate spatial and spectral information, limiting their ability to model the intricate relationships between spectral signatures and spatial structures. 

Despite numerous efforts have been proposed to tackle these challenges, most methods still suffer from two critical problems: 1) \textbf{Static Framework for Feature Extraction.} Existing methods typically use rigid, human-designed frameworks to extract multi-source features, where the computation path is static and fixed. Regardless of the complexity or characteristics of the input data, all features are processed through the same path. This approach fails to account for the varying semantic complexities of different ground objects, causing interference between regions with different levels of detail. As a result, the performance of static frameworks is limited in capturing the diverse nature of HSI and LiDAR data. A dynamic routing mechanism, which adapts the computational path based on the complexity of the input data, is essential to address this shortcoming. 2) \textbf{Hand-crafted cross-modal fusion patterns.} Current feature fusion patterns for HSI and LiDAR classification are mainly hand-crafted and highly dependent on expert knowledge. While these manually designed patterns have achieved success, they may overlook optimal cross-modal interactions, limiting the potential for more effective feature fusion. Furthermore, relying on fixed patterns constrains the ability to generalize across diverse datasets and environments, reducing flexibility and scalability.

To overcome the above-mentioned issues, we propose \textbf{D}ynamic \textbf{C}ross-\textbf{M}odal feature interaction \textbf{Net}work (\textbf{DCMNet}), which is the first multi-source remote sensing image classification framework with dynamic routing mechanism. As illustrated in Fig. \ref{fig_comp}, DCMNet introduces three novel feature interaction blocks: the Bilinear Spatial Attention Block (BSAB), the Bilinear Channel Attention Block (BCAB), and the Integration Convolutional Block (ICB). Each block is designed to capture different aspects of feature interactions between HSI and LiDAR data. The BSAB enhances spatial feature correlations, while the BCAB explicitly models relationships between spectral information from HSI and structural information from LiDAR. Additionally, the ICB is designed to efficiently extract discriminative features for `easy' samples by bypassing complex operations. Importantly, our dynamic routing mechanism determines the optimal computational path for each block based on the input data, enabling the model to adaptively exploit the complementary features of HSI and LiDAR data. By conducting experiments on three HSI and LiDAR datasets, we validate that our proposed DCMNet is superior to other state-of-the-art competitors on both overall performance comparison and visual analyses.

The main contributions of this work can be summarized as follows:

\begin{itemize}
    \item We propose cross-modal feature interaction framework based on dynamic routing for HSI and LiDAR data classification, which contains multiple layers of dynamic blocks for feature interaction exploration. To the best of our knowledge, it is the first work to extract complementary information between multi-source remote sensing data via dynamic routing mechanism.
    
    \item We design three feature interactive blocks, dubbed as BSAB, BCAB and ICB, which are used to capture spatial, spectral, and discriminative features, respectively. Dynamic routing gate is designed in each block to learn different path for semantic diversity.
    
    \item Extensive experiments on three benchmark datasets show that our proposed method achieves competitive performance with several state-of-the-art approaches. As a side contribution, we have released the codes to benefit other researchers in the multi-source remote sensing community.
\end{itemize}

The remainder of this paper is organized as follows: In Section II, we review closely related methods for HSI and LiDAR data classification. The details of DCMNet are introduced in Section III. Experiments on three datasets are shown in Section IV. Conclusions are drawn in Section V.

\section{Related Works}

\subsection{Handcrafted Multi-Source Data Fusion Methods} 

Decision-level fusion \cite{liao2014combining} \cite{bigdeli15} \cite{jia21decision}  and feature-level fusion \cite{chen2017deep} \cite{xu2017multisource}  \cite{li2020a3clnn} have been designed for multi-source remote sensing data fusion. These methods use handcrafted framework for classification. To be specific, decision-level fusion methods use several classifiers individually, and then combine the classification together. Liao et al. \cite{liao2014combining} use four SVM classifiers by using spectral, spatial, elevation, and graph fused features individually as input. Then, the results of four classifiers are combined via weighted voting. Bigdeli et al. \cite{bigdeli15} presented a fuzzy multiple classifier system for multi-source data fusion. fuzzy $k$-NN and fuzzy maximum likelihood are employed as classifiers. Jia et al. \cite{jia21decision} use 2D/3D Gabor filters and superpixel-guided kernel PCA for feature extraction, and three random forest classifiers are employed on three feature cubes. Weighted voting-based decision fusion strategy is incorporated for fusion.  Zhang et al. \cite{zhang2014ensemble} proposed  an ensemble multiple kernel active learning system based on the maximum disagreement query strategy, which integrated the classification results from multiple feature cubes. 

Besides decision-level fusion methods, feature-level fusion methods are more frequently used in multi-source data fusion. Most feature-level fusion methods use dual-branch networks to extract features from multi-source data, and then fuse the extracted features via summation, concatenation, or attention mechanism. Li et al. \cite{li2020a3clnn} presented a spatial, spectral and multi-scale attention LSTM network for multi-source data feature extraction. Dong et al. \cite{dong22tgrs} proposed a multi-branch feature fusion network with self-attention and cross-guided attention for HSI and LiDAR data classification. Li et al. \cite{graph22tgrs} employed a graph-based feature selective network for multi-source image classification. Gao et al. \cite{gao2022tgrs} presented a cross-attention module to extract feature correlations from multi-source data. Li et al. \cite{li2023mixing} proposed a unified framework, mixing self-attention and convolution network (MACN), for comprehensive feature extraction and efficient feature fusion. Gao et al. \cite{gao2024relationship} proposed Relationship Learning From Multisource Images via Spatial-Spectral Perception Network (S2PNet), which leverages spatial perception and spectral perception networks to target the advantages of different data sources while introducing a memory unit to handle cross-information between modalities. Similarly, Zhang et al. \cite{zhang2024remote} presented the Multimodal Adaptive Modulation Network (MAMNet) for landcover classification tasks, employing a cross-modal interacting module (CIM) and a modal attention layer (MAL) for feature representation and fusion.  In addition, contrastive fusion \cite{yang24cf}, anchor graph \cite{cai24ag}, cross-attention mechanism \cite{yang24ca} \cite{roy24ca}, and multi-scale modeling \cite{liu24msmd} \cite{song24hf} \cite{sun24ms}, and Markov edge fusion \cite{wang24markov} are employed for HSI and LiDAR data joint classification.

Recently, advanced network architectures such as Transformers and Mamba-based models have gained popularity in multi-source remote sensing classification. ExViT \cite{yao2023extended}, for instance, extends the Vision Transformer (ViT) with cross-modality attention (CMA) to fuse HSI and LiDAR data. Similarly, SS-MAE \cite{lin2023ss} leverages masked image modeling (MIM) with spatial and spectral branches to jointly process HSI and LiDAR/SAR data, combining global Transformer-based features with local CNN-based extraction for superior classification results. TCPSNet \cite{zhou2024tcpsnet} combines Transformers and cross-pseudo-Siamese learning networks for land use and land cover (LULC) classification, effectively leveraging feature-level and decision-level fusion for robust integration of local and global features. This hybrid approach demonstrates superior performance across various datasets, including Trento and Houston 2013. Additionally, MSFMamba \cite{gao2024msfmamba} builds on the State Space Model (SSM) and incorporates multi-scale strategies to enhance efficiency and address feature redundancy, demonstrating excellent performance on multiple multi-source datasets.

Although the above-mentioned methods have achieved great progress by considering different fine-grained cross-modal feature interactions, their network architectures are essentially handcrafted and heavily depend on expert knowledge. These handcrafted approaches are static in nature, meaning they require predefined computation paths that treat all input data in the same way, regardless of the complexity or characteristics of the data. This static design is not well-suited to the varying complexities of multi-source data like HSI and LiDAR, where some samples may be simpler and others more challenging. In doing so, DCMNet overcomes the limitations of static, handcrafted approaches by introducing a flexible, data-driven framework capable of fully leveraging the complementary strengths of HSI and LiDAR data.

\begin{figure*}[ht]
\centering
\includegraphics [width=6.5in]{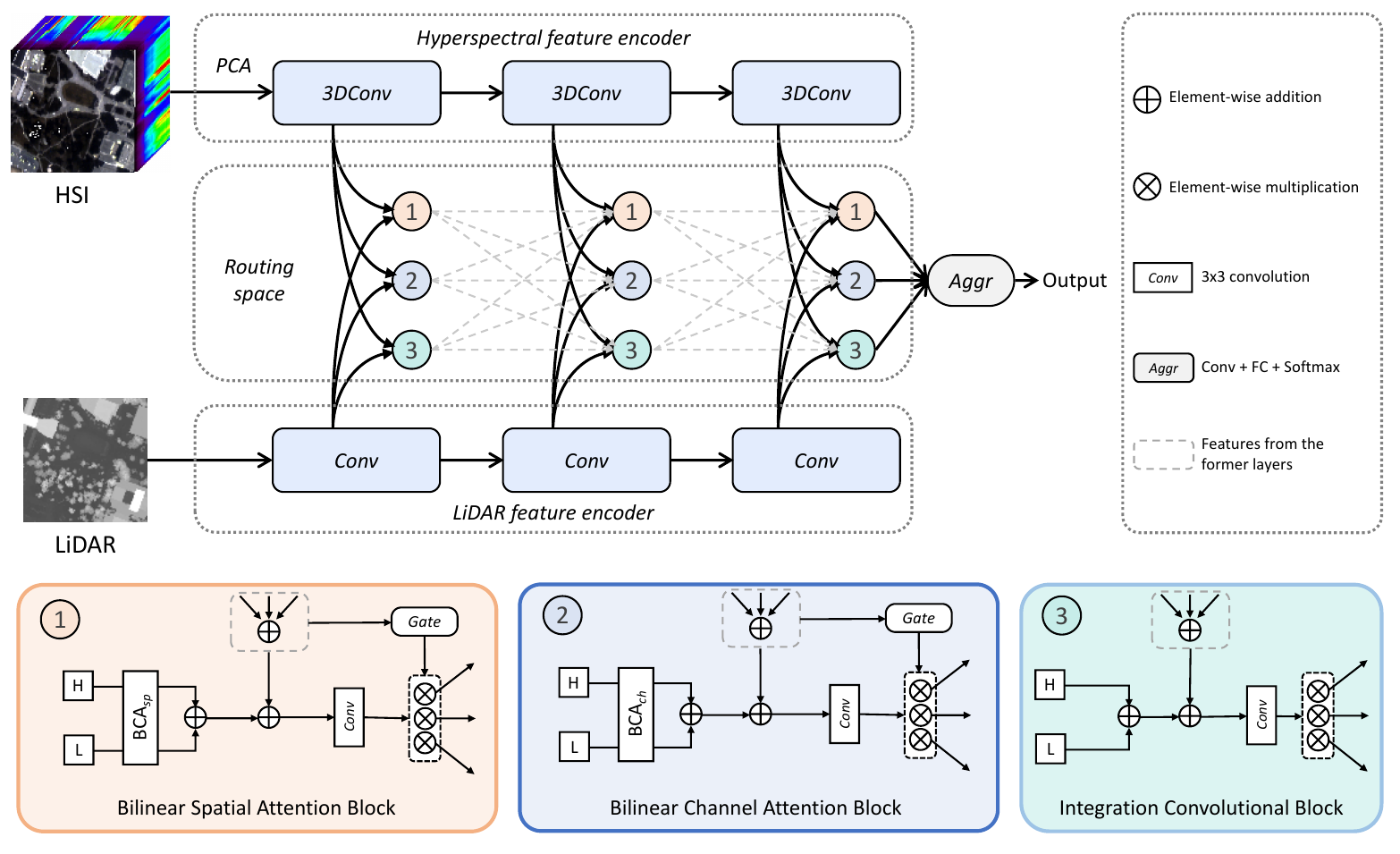}
\caption{The proposed Dynamic Cross-Modal Feature Interaction Network (DCMNet). It is comprised of hyperspectral feature encoder, LiDAR feature encoder, and routing space. In the 3-layer routing space, we design three feature interactive blocks and connect them in a fully connected manner. The details of each block are shown in the rounded rectangle of the corresponding color.}
\label{fig_frame}
\end{figure*}

\subsection{Dynamic Network}

Dynamic networks \cite{huang2018multi} \cite{bolukbasi2017adaptive} \cite{wang2018skipnet} \cite{veit2018convolutional} \cite{sui2024diffusion} generate different execution paths conditioned on the input samples, and have become a hot topic. In dynamic networks, each sample is assigned appropriate computation. The redundant computation of ``easy" samples is reduced, while the computation of ``hard" samples is preserved to extract representative features.

Layer skipping \cite{wang2018skipnet} \cite{veit2018convolutional}, channel pruning \cite{gao2018dynamic} and dynamic filter \cite{chen2019you} are designed to quickly discriminate the simple samples while sustaining the capability of distinguishing difficult ones. Specifically, Wang et al. \cite{wang2018skipnet} designed a gating network to selectively skip convolutional layer based on the activation of the previous layer. Gao et al. \cite{gao2018dynamic} proposed a feature boosting method that amplifys salient convolutional channels and skips unimportant ones at run-time. Chen et al. \cite{chen2019you} investigated the dynamic filter selection in CNNs, and designed a global gated network to generate binary gates to amplify salient filters. Some recent approaches dynamically adjust the convolutional filter neighborhoods according to the input samples, such as CondConv \cite{condconv}, DynamicConv \cite{dynamicconv} and decoupled dynamic filter \cite{zhou21cvpr}. 

Recently, dynamic routing has been incorporated to design dynamic network architecture to the corresponding input. Specifically, a super network with gated units is established first, and the network performs dynamic routing to generate data-dependent paths at run-time. Li et al. \cite{li2020learning} proposed a dynamic routing network, adapting to the object distributions in each image. A soft conditional gate is designed to select different paths on the fly. Qu et al. \cite{qu21sigir} presented a text-image retrieval network based on routing mechanism. Four types of cells are connected in a dense strategy to construct a routing space. Yang et al. \cite{yang22tgrs} proposed a hidden path selection network for remote sensing image segmentation. An extra mini-branch is designed to generate soft masks for pixel-wise path selection. 

To the best of our knowledge, the dynamic routing mechanism has rarely been considered in multi-source remote sensing data classification. Different from related works, our method is the first one to introduce the dynamic routing mechanism for HSI and LiDAR data joint classification.

\begin{table*}[htbp]
\centering
\caption{Detailed parameter configuration for the DCMNet architecture for Houston 2013 dataset.}
\renewcommand{\arraystretch}{1.2}
\begin{tabular}{c|ccc}
\hline\toprule
\textbf{Layer Name} & \textbf{Input Size} & \textbf{Output Size} & \textbf{Kernel Size} \\ \midrule
PCA (HSI Preprocessing) & HSI: (144, 11, 11) & (30, 11, 11) & PCA projection \\ 
3D Conv1 (HSI Feature Extraction) & (1, 30, 11, 11) & (8, 22, 9, 9) & 3D CNN (9x3x3, stride=1, padding=0) \\ 
3D Conv2 (HSI Feature Extraction) & (8, 22, 9, 9) & (16, 16, 7, 7) & 3D CNN (7x3x3, stride=1, padding=0) \\ 
3D Conv3 (HSI Feature Extraction) & (16, 16, 7, 7) & (32, 12, 5, 5) & 3D CNN (5x3x3, stride=1, padding=0) \\ 
2D Conv1 (LiDAR Feature Extraction) & LiDAR: (1, 11, 11) & (64, 9, 9) & 2D CNN (3x3, stride=1, padding=0) \\ 
2D Conv2 (LiDAR Feature Extraction) & (64, 9, 9) & (128, 7, 7) & 2D CNN (3x3, stride=1, padding=0) \\ 
2D Conv3 (LiDAR Feature Extraction) & (128, 7, 7) & (128, 3, 3) & 2D CNN (5x5, stride=1, padding=0) \\ 
Projector (Input to Routing Space) & (C, P, P) & (128, 3, 3) & 2D CNN ((P-2)x(P-2), stride=1, padding=0) \\ 
Router (Feature Interactive Blocks Routing) & (128, 3, 3) & (256, 256) & N/A \\
BSAB (Bilinear Spatial Attention Block) & (128, 3, 3), (128, 3, 3) & (128, 3, 3) & N/A \\ 
BCAB (Bilinear Channel Attention Block) & (128, 3, 3), (128, 3, 3) & (128, 3, 3) & N/A \\ 
ICB (Integration Convolutional Block) & (128, 3, 3), (128, 3, 3) & (128, 3, 3) & N/A \\
Aggregation Layer (Final Output) & (128, 3, 3) & 15 & N/A \\ \bottomrule\hline
\end{tabular}
\label{tab:dcmnet_structure}
\end{table*}

\section{Methodology}

\subsection{Overall Framework of the Proposed DCMNet}

The overall architecture of our DCMNet is shown in Fig. \ref{fig_frame}. It is comprised of hyperspectral feature encoder, LiDAR feature encoder, and the routing space. Specifically, given a pair of coregistered HSI patch $I_h\in\mathbb{R}^{c\times p \times p}$ and LiDAR patch $I_l\in\mathbb{R}^{1\times p \times p}$ that cover the same region, the aim of joint classification is to assign label to the region. Here $c$ denotes the number of spectral bands in HSI, and $p$ denotes the patch size of the input data. First, PCA is employed to reduce the spectral dimension of HSI to retain the most important information, and $K$ components are retained. Afterwards, two feature encoders with three convolutional layers are applied to obtain three-level features from HSI and LiDAR data. Therefore, the obtained three-level HSI features can be denoted as $\{\mathbf{F}^1_h; ~ \mathbf{F}^2_h; ~ \mathbf{F}^3_h\}$, and the three-level LiDAR features can be denoted as $\{\mathbf{F}^1_l; ~ \mathbf{F}^2_l; ~ \mathbf{F}^3_l\}$. These features are fed into the routing space to dynamically exploit the complementary features between HSI and LiDAR data.

We construct a 3-layer routing space to capture data-dependent feature extraction paths via dynamic routing mechanism. In the routing space, we design three feature interactive blocks: Bilinear Spatial Attention Block (BSAB), Bilinear Channel Attention Block (BCAB), and Integration Convolutional Block (ICB). We deploy BSAB, BCAB, and ICB in parallel at each layer, and build a 3-layer routing space in fully connected manner. Each block contains a calculation unit and a routing gate. The calculation unit is in charge of feature extraction, while the routing gate is responsible for selecting the direction of the output signal.

\begin{figure*}[h]
\centering
\includegraphics [width=7in]{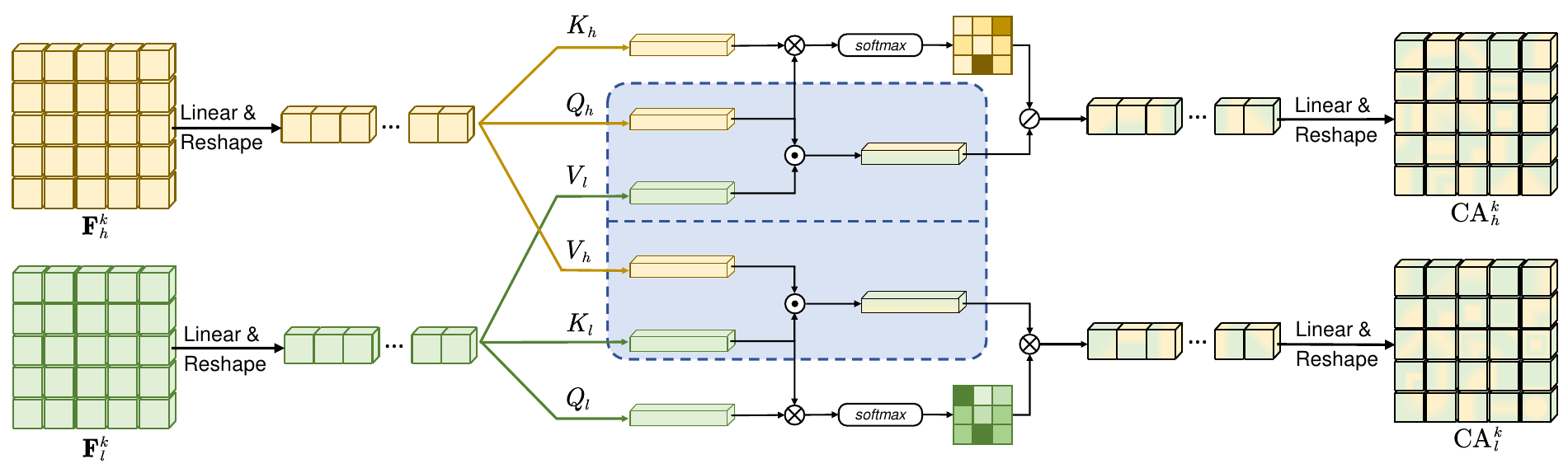}
\caption{Illustration of the bilinear cross-attention. Second-order feature interactions are employed for feature extraction. In BSAB, the attention matrix between query and value is computed along the spatial dimension. In BCAB, the attention matrix between query and value is computed along the channel dimension.}
\label{fig_bca}
\end{figure*}

For simplicity, BSAB, BCAB and ICB are numbered as \textbf{1}, \textbf{2}, and \textbf{3}, respectively. We define the calculation unit as:
\begin{equation}
\mathbf{H}^k_i= E^k_i ~ (\mathbf{X}^k_i, \mathbf{F}^k_h, \mathbf{F}^k_l),
\end{equation}
where $\mathbf{X}^k_i$ denotes the input of the $i$-th ($1\leq i \leq 3$) block in the $k$-th layer ($1\leq k \leq 3$), and $\mathbf{H}^k_i$ denotes the output of the $i$-th block in the $k$-th layer. $\mathbf{F}^k_h$ denotes the features from HSI encoder in the $k$-th layer, and $\mathbf{F}^k_l$ denotes the features from LiDAR encoder in the $k$-th layer.

In the routing space, the HSI and LiDAR features would obtain the optimal combination by using the dynamic routing mechanism. Finally, the fused features with rich complementary information are fed into fully connected layers to compute the classification results.

To provide a comprehensive understanding of the DCMNet architecture and its detailed configurations for Houston 2013 dataset, we present Table \ref{tab:dcmnet_structure}, whcih includes the input and output sizes of features for each layer and the size of the convolution kernels used in both the HSI and LiDAR feature extraction processes. It is important to note that since the feature dimensions produced by the HSI and LiDAR encoders differ at each layer, we employ a projector consisting of 2D convolution layers before feeding the features into the dynamic routing space. This projector ensures that the features from the two encoders are mapped into a unified space, allowing them to be processed together within feature interactive blocks in the routing space. By standardizing the feature dimensions, the projector enables efficient feature integration, which is critical for the dynamic routing mechanism to operate effectively. For simplicity in our formula descriptions, we omit the explicit notation for the projector and represent the features after projection with $\mathbf{F}$.

\subsection{Bilinear Spatial Attention Block (BSAB)}

The HSI and LiDAR features derived from the encoders commonly contain a large amount of spatial information redundancy. To reduce the spatial feature redundancy and enable the network to focus on the critical features, we compute the spatial-wise bilinear cross-attention $\textrm{BCA}_{sp}(\cdot)$ in the $k$-th layer as illustrated in Fig. \ref{fig_bca} as follows:
\begin{equation}
\mathbf{CA}^k_h, \mathbf{CA}^k_l = \textrm{BCA}_{sp} ~ (\mathbf{F}^k_h, \mathbf{F}^k_l),
\end{equation}
where $\mathbf{F}^k_h\in\mathbb{R}^{c\times p\times p}$ and $\mathbf{F}^k_l \in\mathbb{R}^{c\times p\times p}$ denote features from the HSI encoder and LiDAR encoder, respectively. $c$ is the channel number of the input features, and $p$ is the spatial size of the input features. 

For BCA$_{sp}$ computation, $\mathbf{F}^k_h$ from HSI encoder is fed into the linear layer to obtain $Q_h$, $K_h$, and $V_h$ with the size of $c \times d$, where $d$ denotes the feature dimension. Similarly, we obtain $Q_l$, $K_l$, and $V_l$ from the LiDAR encoder. Next, $V_h$ and $Q_l$ are combined via element-wise multiplication, while $V_l$ and $Q_h$ are combined in the same way as follows:
\begin{equation}
B_h = V_h \odot Q_l,
\end{equation}
\begin{equation}
B_l = V_l \odot Q_h.
\end{equation}

Subsequently, taking $B_h$ and $B_l$ as values, we then compute cross-attention as follows:
\begin{equation}
\textrm{CA}^k_h=\textrm{Attn}_{sp}(Q_h, K_h, B_h) =\textrm{softmax} \left(\frac{Q_h K^T_h}{\sqrt{d}} \right) B_h,
\end{equation}
\begin{equation}
\textrm{CA}^k_l=\textrm{Attn}_{sp}(Q_l, K_l, B_l) =\textrm{softmax} \left(\frac{Q_l K^T_l}{\sqrt{d}} \right) B_l. ~~
\end{equation}
It should be noted that in BCA$_{sp}$, the attention matrix between query and value is computed along the spatial dimension. 

The calculation unit $E^k_1(\cdot)$ of our BSAB is defined as:
\begin{equation}
\mathbf{H}^k_1= \textrm{Conv}(\mathbf{CA}^k_h+\mathbf{CA}^k_l+\mathbf{X}^k_1),
\end{equation}
where $\textrm{Conv}(\cdot)$ denotes the $3\times3$ convolution. In the first layer ($k=1$), since there are no input features from the former layer, the input feature $\mathbf{X}^1_1$ is null.

\subsection{Bilinear Channel Attention Block (BCAB)}

For HSI feature extraction, the spectral information is important that may affect the final classification results. To fully exploit the correlations between the spectral information from HSI and structural information from LiDAR, we compute the channel-wise bilinear cross-attention BCA$_{ch}(\cdot)$ in the $k$-th layer as follows:
\begin{equation}
\mathbf{CA}^k_h, \mathbf{CA}^k_l = \textrm{BCA}_{ch} ~ (\mathbf{F}^k_h, \mathbf{F}^k_l).
\end{equation}

Similarly to $\textrm{BCA}_{sp}$, $\mathbf{F}^k_h$ is fed into a linear layer to obtain $Q_h$, $K_h$, and $V_h$ with the size of $c\times d$. Here $c$ is the channel numbers and $d$ is the feature dimension. $\mathbf{F}^k_l$ is fed into a linear layer to obtain $Q_l$, $K_l$, and $V_l$. Next, $B_h$ is computed via element-wise multiplication between $V_h$ and $Q_l$, while $B_l$ is computed via element-wise multiplication between $V_l$ and $Q_h$.

Next, we compute cross-attention as follows:
\begin{equation}
\textrm{CA}^k_h=\textrm{Attn}_{ch}(Q_h, K_h, B_h) =\textrm{softmax} \left(\frac{Q_h K^T_h}{\sqrt{d}} \right) B_h,
\end{equation}
\begin{equation}
\textrm{CA}^k_l=\textrm{Attn}_{ch}(Q_l, K_l, B_l) =\textrm{softmax} \left(\frac{Q_l K^T_l}{\sqrt{d}} \right) B_l. ~~
\end{equation}
It should be noted that in BCA$_{ch}$, the attention matrix between query and value is computed along the channel dimension.

The calculation unit $E^k_1(\cdot)$ of BCAB is as follows:
\begin{equation}
\mathbf{H}^k_2= \textrm{Conv}(\mathbf{CA}^k_h+\mathbf{CA}^k_l+\mathbf{X}^k_2),
\end{equation}
where $\textrm{Conv}(\cdot)$ denotes the $3\times3$ convolution. In the first layer ($k=1$), since there are no input features from the former layer, the input feature $\mathbf{X}^1_2$ is null.

\subsection{Integration Convolutional Block (ICB)}

Humans can easily understand a simple image. We argue that complicated feature interactions may not always be necessary for `easy' samples. Motivated by this, we present an Integration Convolutional Block (ICB) that could skip unnecessary operations and retain discriminative information. The ICB is formulated as:

\begin{equation}
\mathbf{H}^k_3=\textrm{Conv}(\mathbf{F}^k_h + \mathbf{F}^k_l+\mathbf{X}^k_3)
\end{equation}

\subsection{Dynamic Routing}

To achieve data-aware dynamic routing, we fully connect the cells of adjacent layers and then add a routing gate in each cell. Therefore, we build a multi-layer fully connected routing space. Each block selectively receive signals from the previous layer, and also selectively pass signals to the next layer. Hence, a dynamic routing space is built to explore potential feature interactive patterns. 

\textbf{Soft Routing Gate.} There is a routing gate in each block. 
Each block contains a routing gate, which generates the the probability of passing signals to each block in the next layer based on the input data. To be specific, the routing gate computes the path probability $W$ as follows:
\begin{equation}
W^k_i= \delta (\textrm{FC}(\textrm{ReLU}(\textrm{FC}(\mathbf{F}_h + \mathbf{F}_l+\mathbf{X}^k_i)))),
\end{equation}
where $\delta(\cdot)$ is the gating function. $\mathbf{F}_h$ and $\mathbf{F}_l$ denote features from the HSI and LiDAR encoder, respectively. FC represents the fully connected layer. The restricted Tanh is employed as the gating activation. The gating function is defined as follows:
\begin{equation}
\delta(\cdot) = \max(0, \textrm{Tanh}(\cdot)).
\end{equation}
In each block, the routing gate $W^k_i=\{ w^k_{i,1}, w^k_{i,2}, w^k_{i,3}\}$ is generated. $w^k_{i,j}$ denotes the path probability from the $i$-th block in the $k$-th layer to the $j$-th block in the $(k+1)$-th layer.

\textbf{Routing Process.} Each block contains a calculation unit and a routing gate that performs the path decisions. The routing gate is employed on the output of the calculation unit to generate the input of the next layer. Specifically, the input of the $i$-th block in the ($k+1$)-th layer is obtained by the following aggregation operation:
\begin{equation}
\mathbf{X}^{k+1}_i = \sum^3_{j=1}w^k_{i,j}\mathbf{H}^k_j,
\end{equation}
where $g^k_{j,i}$ is the path probability from the $j$-th block in the $k$-th layer to the $i$-th block in the ($k+1$)-th layer.

Finally, the outputs of the three feature interactive blocks are aggregated at the third layer to compute the classification result.

\definecolor{m1}{HTML}{002FFF}
\definecolor{m2}{HTML}{4EC8ED}
\definecolor{m3}{HTML}{99CD6B}
\definecolor{m4}{HTML}{FED20D}
\definecolor{m5}{HTML}{EE3424}
\definecolor{m6}{HTML}{7D1416}

\begin{table}[htb]
\centering
\caption{The number of training and testing samples for Trento dataset.}
\renewcommand{\arraystretch}{1.2}
\begin{tabular}{cccc|cc}
\hline\toprule
    No. & Name & Color & ~ & Train & Test \\
\midrule
    1 & Apple trees & \cellcolor{m1} & & 129 & 4034\\
    2 & Buildings & \cellcolor{m2} & & 125 & 2903\\
    3 & Ground & \cellcolor{m3} & & 105 & 479\\
    4 & Woods & \cellcolor{m4} & & 154 & 9123\\
    5 & Vineyard & \cellcolor{m5} & & 184 & 10501\\
    6 & Roads & \cellcolor{m6} & & 122 & 3174\\
\midrule
    \multicolumn{3}{c}{Total} & & 819 & 30214\\
\bottomrule\hline
\end{tabular}
 \label{table_trento}
\end{table}

\definecolor{h1}{HTML}{000083}  
\definecolor{h2}{HTML}{0000CB}  
\definecolor{h3}{HTML}{0013FF}  
\definecolor{h4}{HTML}{005BFF}  
\definecolor{h5}{HTML}{00A7FF}   
\definecolor{h6}{HTML}{00EFFF}  
\definecolor{h7}{HTML}{37FFC7}  
\definecolor{h8}{HTML}{83FF7B}  
\definecolor{h9}{HTML}{CBFF33}  
\definecolor{h10}{HTML}{FFEB00} 
\definecolor{h11}{HTML}{FFA300} 
\definecolor{h12}{HTML}{FF5700} 
\definecolor{h13}{HTML}{FF0F00} 
\definecolor{h14}{HTML}{C70000}    
\definecolor{h15}{HTML}{7F0000} 

\begin{table}[htb]
\centering
\caption{The number of training and testing samples for Houston 2013 dataset.}
\renewcommand{\arraystretch}{1.2}
\begin{tabular}{cccc|cc}
\hline\toprule
    No. & Name & Color & ~ & Train & Test \\
\midrule
    1 & Health grass & \cellcolor{h1} & & 198 & 1053\\
    2 & Stressed grass & \cellcolor{h2} & & 190 & 1064\\
    3 & Synthetic grass & \cellcolor{h3} & & 192 & 505\\
    4 & Trees & \cellcolor{h4} & & 188 & 1056\\
    5 & Soil & \cellcolor{h5} & & 186 & 1056\\
    6 & Water & \cellcolor{h6} & & 182 & 143\\
    7 & Residential & \cellcolor{h7} & & 196 & 1072\\
    8 & Commercial & \cellcolor{h8} & & 191 & 1053\\
    9 & Road & \cellcolor{h9} & & 193 & 1059\\
    10 & Highway & \cellcolor{h10} & & 191 & 1036\\
    11 & Railway & \cellcolor{h11} & & 181 & 1054\\
    12 & Parking lot 1 & \cellcolor{h12} & & 192 & 1041\\
    13 & Parking lot 2 & \cellcolor{h13} & & 184 & 285\\
    14 & Tennis court & \cellcolor{h14} & & 181 & 247\\
    15 & Running track & \cellcolor{h15} & & 187 & 473\\
\midrule
    \multicolumn{4}{c}{Total} & 2832 & 12197\\
\bottomrule\hline
\end{tabular}
 \label{table_Houston}
\end{table}

\definecolor{hh1}{HTML}{32CD33} 
\definecolor{hh2}{HTML}{ADFF30} 
\definecolor{hh3}{HTML}{008081}  
\definecolor{hh4}{HTML}{228B22}  
\definecolor{hh5}{HTML}{2E4F4E}  
\definecolor{hh6}{HTML}{8B4512} 
\definecolor{hh7}{HTML}{00FFFF}
\definecolor{hh8}{HTML}{FFFFFF}
\definecolor{hh9}{HTML}{D3D3D3}
\definecolor{hh10}{HTML}{FE0000}
\definecolor{hh11}{HTML}{A9A9A9}
\definecolor{hh12}{HTML}{696969}
\definecolor{hh13}{HTML}{8B0001}
\definecolor{hh14}{HTML}{C86400}
\definecolor{hh15}{HTML}{FEA500}
\definecolor{hh16}{HTML}{FFFF00}
\definecolor{hh17}{HTML}{DAA521}
\definecolor{hh18}{HTML}{FF00FE}
\definecolor{hh19}{HTML}{0000FE}
\definecolor{hh20}{HTML}{3FE0D0}
 
\begin{table}[htb]
\centering
\caption{The number of training and testing samples for Houston 2018 dataset.}
\renewcommand{\arraystretch}{1.2}
\begin{tabular}{cccc|cc} 
\hline\toprule
    No. & Name & Color & ~ & Train & Test \\
\midrule
1 & Health grass & \cellcolor{hh1} & & 1000 & 38196\\
2 & Stressed grass & \cellcolor{hh2} & & 1000 & 129008\\
3 & Artificial turf & \cellcolor{hh3} & & 1000 & 1736\\
4 & Evergreen trees & \cellcolor{hh4} & & 1000 & 53322\\
5 & Deciduous trees & \cellcolor{hh5} & & 1000 & 19172\\
6 & Bare earth & \cellcolor{hh6} & & 1000 & 17064\\
7 & Water & \cellcolor{hh7} & & 500 & 564\\
8 & Residential buildings & \cellcolor{hh8} & & 1000 & 157995\\
9 & Non-residential buildings & \cellcolor{hh9} & & 1000 & 893769\\
10 & Roads & \cellcolor{hh10} & & 1000 & 182283\\
11 & Sidewalks & \cellcolor{hh11} & & 1000 & 135035\\
12 & Crosswalks & \cellcolor{hh12} & & 1000 & 5059\\
13 & Major thoroughfares & \cellcolor{hh13} & & 1000 & 184438\\
14 & Highways & \cellcolor{hh14} & & 1000 & 38438\\
15 & Railways & \cellcolor{hh15} & & 1000 & 26748\\
16 & Paved parking lots & \cellcolor{hh16} & & 1000 & 44932\\
17 & Unpaved parking lots & \cellcolor{hh17} & & 250 & 337\\
18 & Cars & \cellcolor{hh18} & & 1000 & 25289\\
19 & Trains & \cellcolor{hh19} & & 1000 & 20479\\
20 & Stadium seats & \cellcolor{hh20} & & 1000 & 26296\\
\midrule
\multicolumn{4}{c}{Total} & 18750 & 2000160\\
\bottomrule\hline
\end{tabular}
 \label{table_Houston2018}
\end{table}

\begin{table}[t]
\centering
\caption{Classification performance in terms of overall accuracy with different feature sizes in the routing space.}
\renewcommand{\arraystretch}{1.2}
\begin{tabular}{c|c|ccc} 
\hline\toprule
$d$ & $c$ & ~~~ Trento ~  & Houston 2013 & Houston 2018 ~\\ 
\midrule
\multirow{4}*{9} & 16 & 95.71 & 93.67 & 89.77 \\ 
~ & 32 & 97.61 & 92.99 & 90.42 \\ 
~ & 64 & 98.13 & 93.96 & 91.08 \\ 
~ & 128 & 98.72 & \textbf{95.11} & 90.91 \\ 
\midrule
\multirow{4}*{25} & 16 & 96.65 & 92.72 & 91.86 \\ 
~ & 32 & \textbf{98.96} & 94.42 &92.88 \\ 
~ & 64 & 98.64 & 93.95 &\textbf{93.27} \\ 
~ & 128 & 97.95 & 93.58 & 92.33 \\
\midrule
\multirow{4}*{49} & 16 & 95.89 & 93.03 &91.53 \\ 
~ & 32 & 96.41 & 93.41 & 92.64 \\ 
~ & 64 & 97.03 & 93.97 &92.97 \\ 
~ & 128 & 96.47 & 93.63 & 91.65 \\ 
\midrule
\multirow{4}*{81} & 16 & 97.86 & 92.95 & 90.47 \\ 
~ & 32 & 96.53 & 92.87 & 91.61\\ 
~ & 64 & 95.94 & 92.44  & 90.39\\ 
~ & 128 & 95.38 & 91.66 & 88.01 \\ 
\bottomrule\hline
\end{tabular}
\label{table_unify}
\end{table}

\begin{table*}[t]
\renewcommand\arraystretch{1.2}
\centering
\caption{Classification performance of different methods on Trento dataset.}
\begin{tabular}{c|cccccccccccc}
\hline\toprule
~~~Class~~~ & FusAtNet & ~TBCNN~ & ~~DFINet~ &  AsyFFNet & ~$S^2$ENet~ & MACN & SS-MAE & MSFMamba & ~DCMNet~\\\hline
Apple trees & 99.15 & 98.98 & 98.98 & 99.62 & 99.92 & 97.37 & 97.90 & 98.36 & 99.38 \\ 
Buildings & 99.21 & 98.56 & 99.96 & 99.57 & 99.96 & 94.64 & 98.02 & 98.09 & 94.39 \\ 
Ground & 100.00 & 88.23 & 91.18 & 85.29 & 93.32 & 85.84 & 93.32 & 81.55 & 93.95 \\
Woods & 99.50 & 99.93 & 99.99 & 99.98 & 99.98 & 99.77 & 99.96 & 98.98 & 100.00 \\
Vineyard & 99.69 & 99.78 & 99.73 & 99.88 & 97.88 & 98.49 & 97.65 & 98.99 & 99.71 \\
Roads & 89.68 & 91.64 & 90.86 & 89.91 & 89.28 & 91.92 & 96.85 & 91.11 & 97.89 \\\hline
OA & 98.49 & 98.61 & 98.70 & 98.63 & 98.01 & 97.53 & 98.28 & 97.78 & 98.96 \\ 
AA & 97.87 & 96.18 & 96.78 & 95.71 & 96.67 & 94.67 & 97.28 & 94.51 & 97.55 \\ 
Kappa & 97.96 & 98.14 & 98.26 & 98.16 & 97.34 & 96.69 & 97.71 & 97.03 & 98.61 \\ 
\bottomrule\hline
\end{tabular}
\label{trento_compare_table}
\end{table*}

\section{Experiments And Analysis}

\subsection{Dataset Description}

To demonstrate the advantages of the proposed DCMNet, we evaluate the network on three well-known HSI and LiDAR datasets: Trento, Houston 2013, and Houston 2018.

\textbf{Trento Dataset.} This dataset was captured at a rural area in the south of the city of Trento, Italy. The AISA Eagle collected HS data consists of 63 spectral bands with wavelength ranging from 0.42 to 0.99 $\mu$m. The LiDAR data were taken by the Optech ALTM 3100EA sensor. Table \ref{table_trento} lists the number of training and testing samples for six classes.

\textbf{Houston 2013 Dataset.} This dataset was obtained in 2012 by an airborne sensor over the University of Houston campus and the surrounding area. The dataset was introduced in the 2013 IEEE GRSS Data Fusion Contest. The hyperspectral data consists of 144 spectral bands with a wavelength ranging from 0.38 to 1.05 $\mu m$, including 15 classes. Table \ref{table_Houston} lists the number of training and testing samples of each class. 

\textbf{Houston 2018 Dataset.} This dataset was acquired in the same location as Houston 2013. It has been made publicly available in the 2018 GRSS Data Fusion Contest. The dataset is comprised of LiDAR, HSI, and optical remote sensing data, with 7, 48, and 3 channels, respectively. The HSI data cover a 380–1050 nm spectral range. The spatial resolution of HSI is 1 m, while the spatial resolution of LiDAR data is 0.5 m. We rescale the HSI to ensure consistency of the multi-source data. Table \ref{table_Houston2018} lists the number of training and testing samples for 20 classes.

\begin{figure}[htb]
\centering
\includegraphics[width=3.1in]{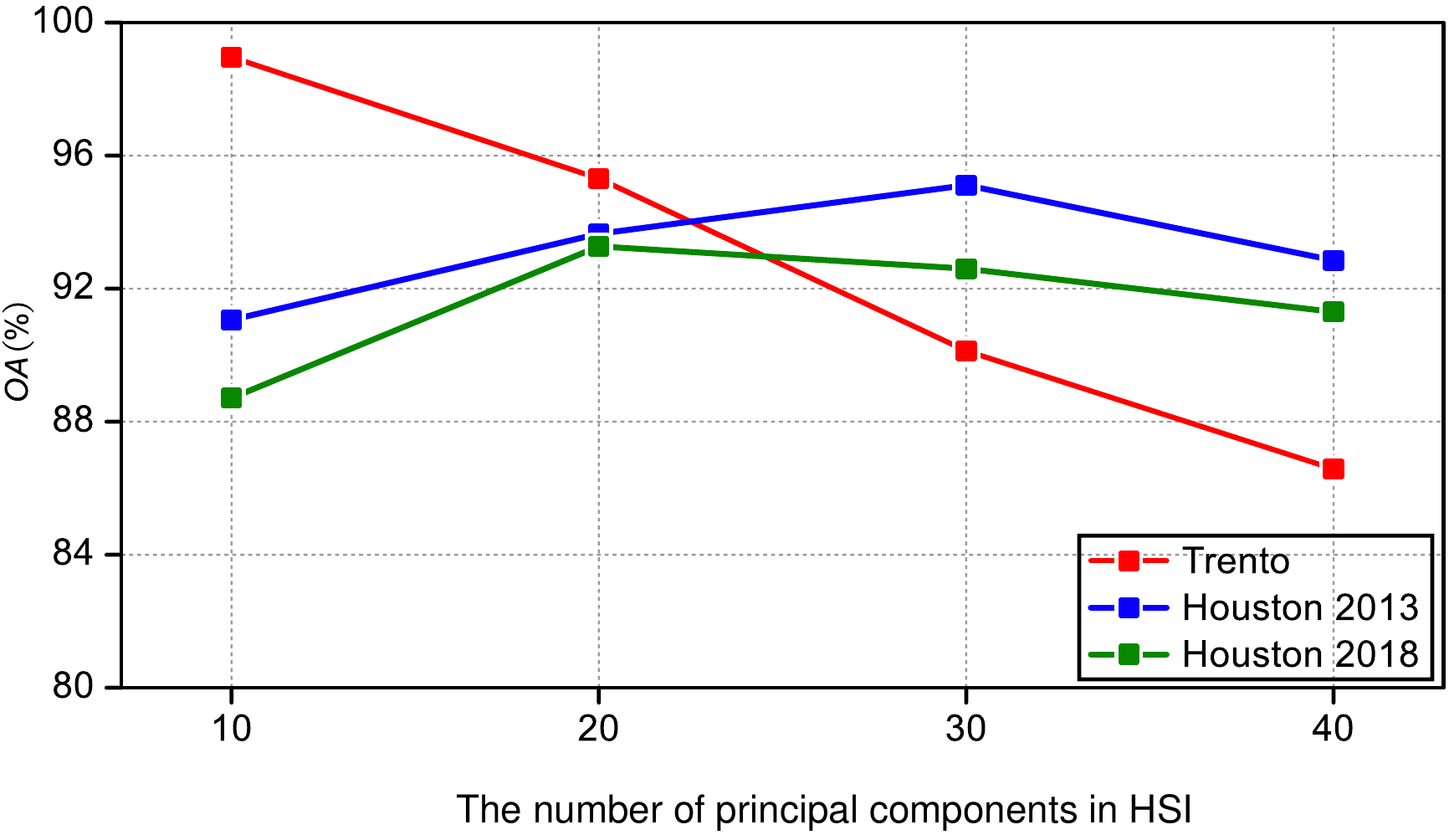}
\caption{The number of principal components in HSI versus the classification accuracy.}
\label{fig_para_k}
\end{figure}

\begin{figure}[htb]
\centering
\includegraphics[width=3.1in]{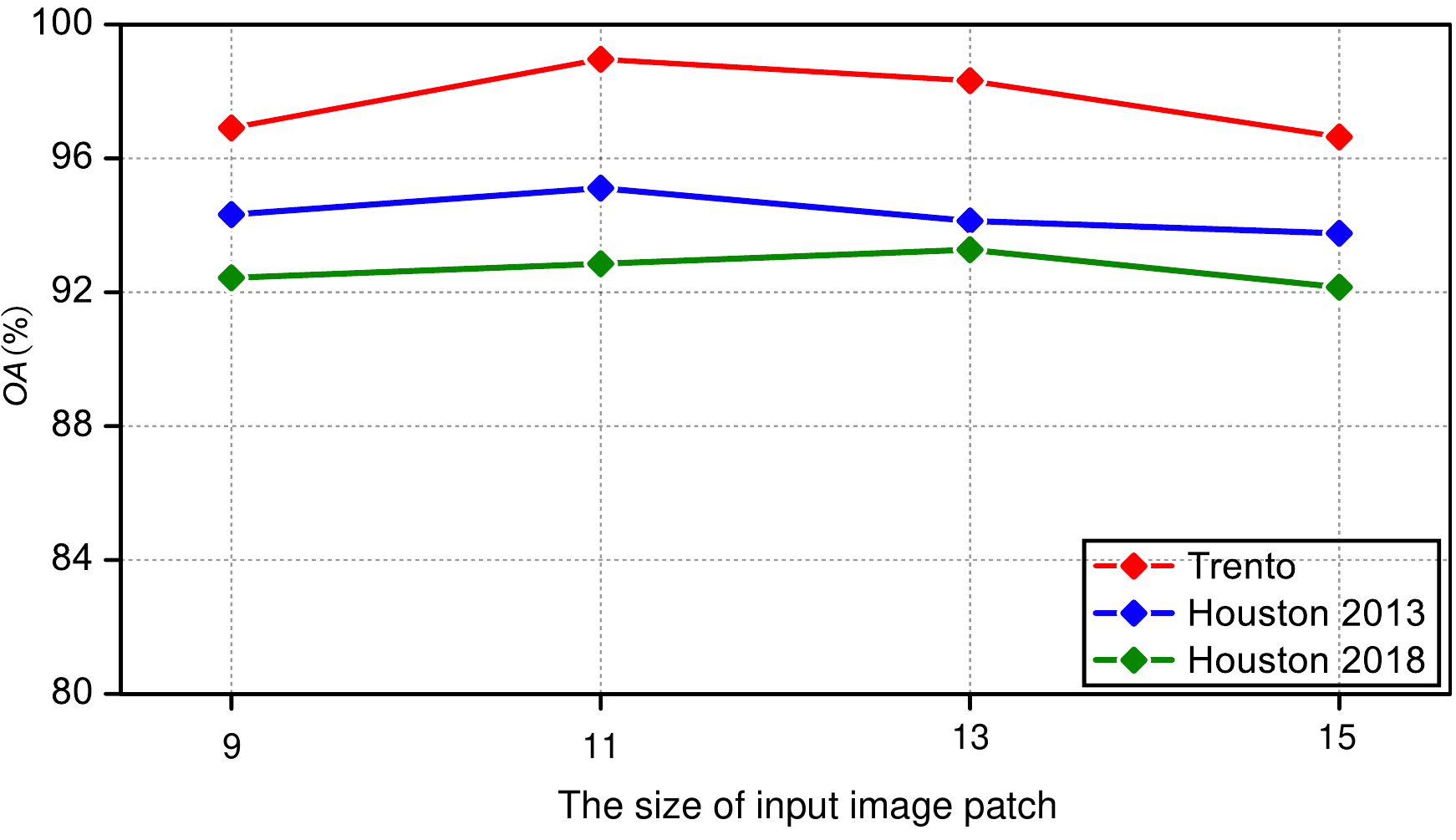}
\caption{The size of input image patch versus the classification accuracy.}
\label{fig_para_p}
\end{figure}

\begin{figure*}[htb]
\centering
\includegraphics [width=0.8\textwidth]{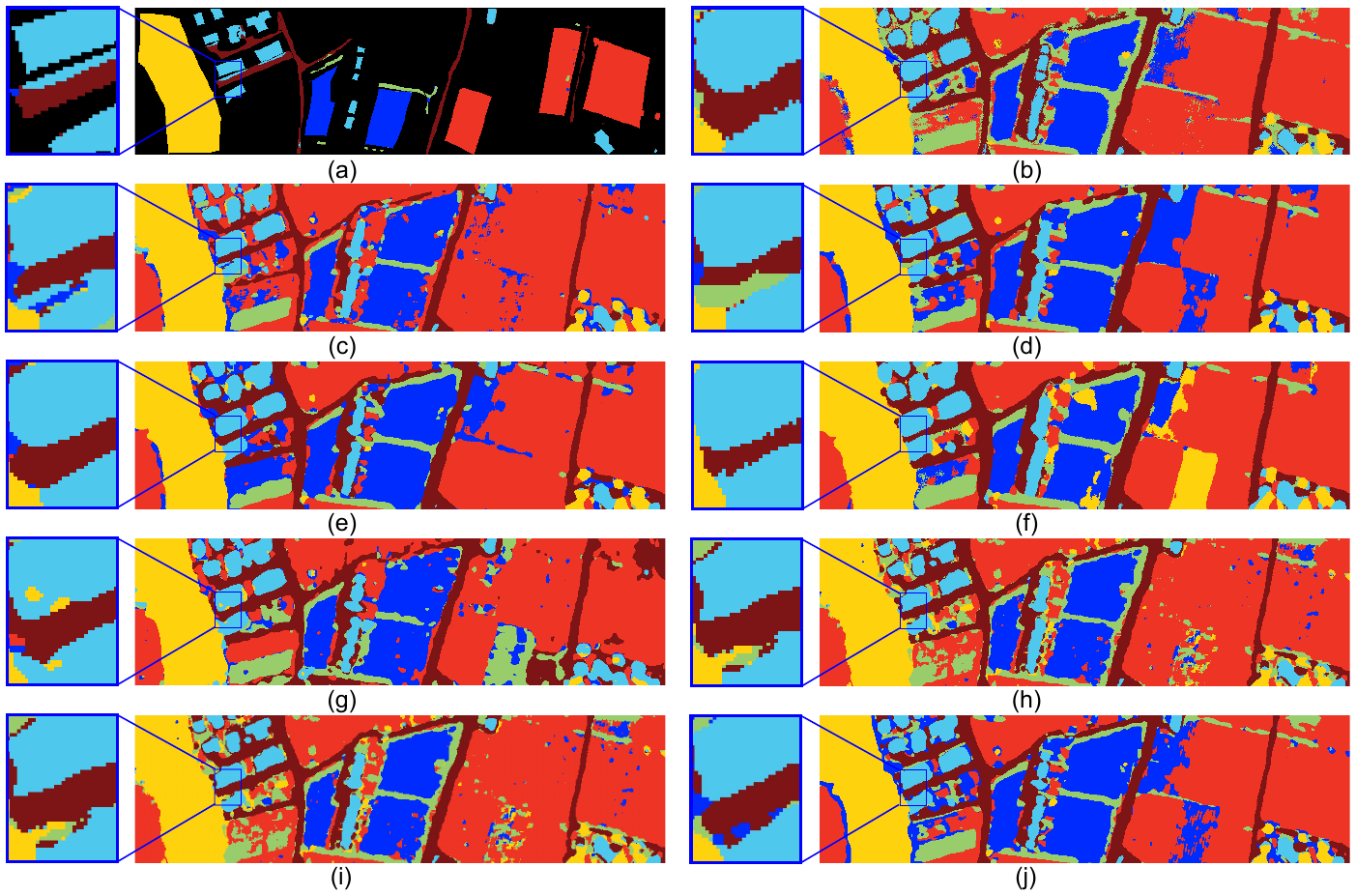}
\caption{Visualized classification maps of different methods on Trento dataset. (a) Ground Truth. (b) FusAtNet. (c) TBCNN. (d) DFINet. (e) AsyFFNet. (f) $S^2$ENet. (g) MACN. (h) SS-MAE. (i) MSFMamba. (j) DCMNet.}
\label{trento_fig_com}
\end{figure*}

\begin{figure*}[htbp]
\centering
\includegraphics [width=0.99\textwidth]{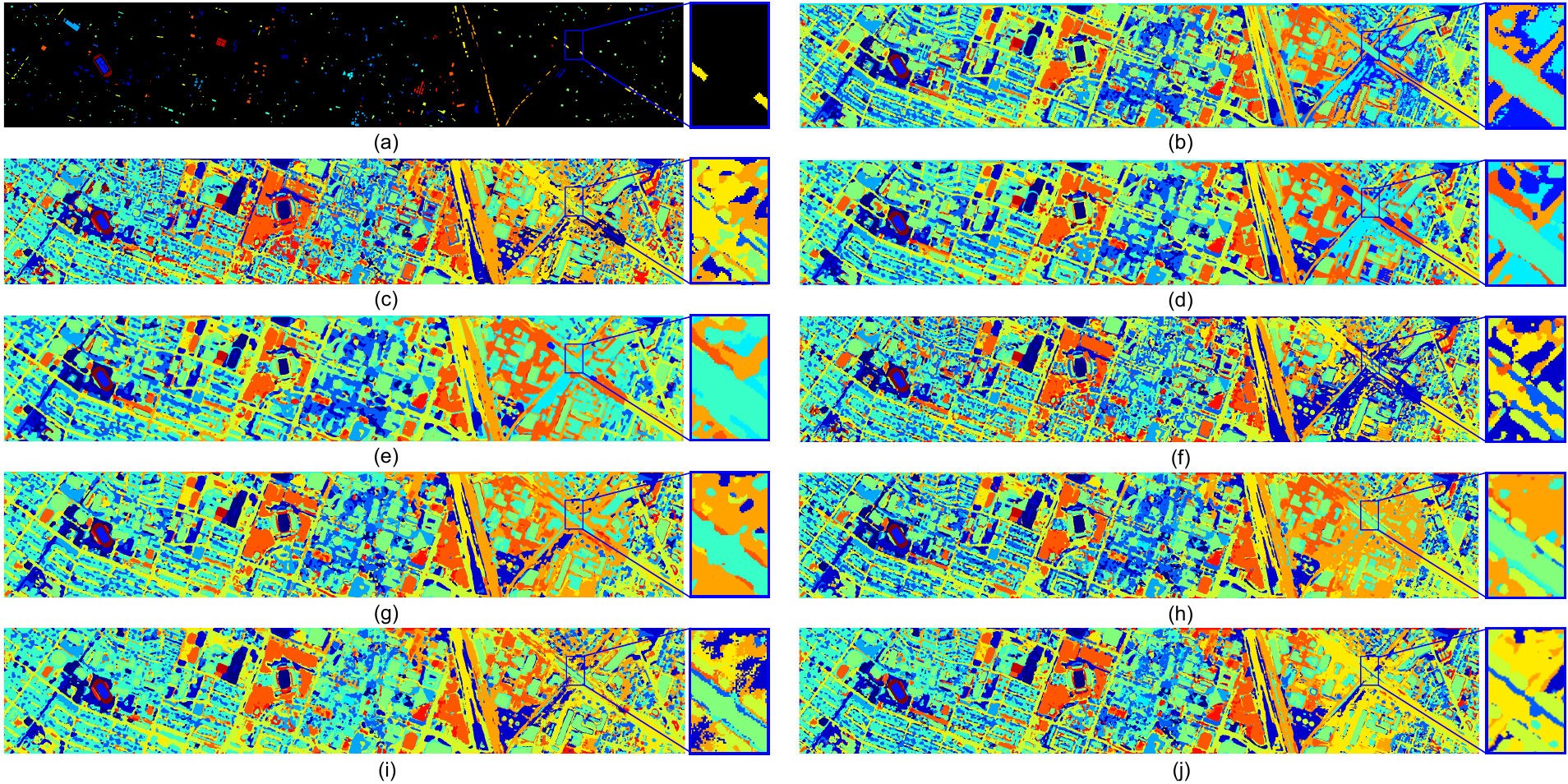}
\caption{Visualized classification maps of different methods on Houston 2013 dataset. (a) Ground Truth. (b) FusAtNet. (c) TBCNN. (d) DFINet. (e) AsyFFNet. (f) $S^2$ENet. (g) MACN. (h) SS-MAE. (i) MSFMamba. (j) DCMNet.}
\label{houston_fig_com}
\end{figure*}

\begin{table*}[htb]
\renewcommand\arraystretch{1.2}
\centering
\caption{Classification performance of different methods on Houston 2013 Dataset.}
\begin{tabular}{c|cccccccccc}
\hline\toprule
~~~Class~~~ & FusAtNet & ~TBCNN~ & ~~DFINet~ &  AsyFFNet & ~$S^2$ENet~ & MACN & SS-MAE & MSFMamba & ~DCMNet~\\\hline
Healthy grass & ~82.34~ & ~80.15~ & ~83.10~ & ~82.43~ & ~83.09~ & 82.91 & 82.91 & 100.00 & ~80.72~ \\ 
Stressed grass & ~84.77~ & ~99.24~ & ~85.15~ & ~85.15~ & ~85.53~ & 100.00 & 85.15 & 86.72 & ~98.78~ \\ 
Synthetic grass & ~100.00~ & ~100.00~ & ~97.82~ & ~99.01~ & ~99.80~ & 99.80 & 100.00 & 100.00 & ~100.00~ \\
Trees & ~93.28~ & ~96.68~ & ~92.05~ & ~92.23~ & ~99.34~ & 99.05 & 97.82 & 93.77 & ~99.53~ \\ 
Soil & ~99.81~ & ~99.52~ & ~99.81~ & ~100.00~ & ~99.91~ & 100.00 & 100.00 & 99.71 & ~99.91~ \\ 
Water & ~100.00~ & ~93.70~ & ~80.42~ & ~95.80~ & ~100.00~ & 95.80 & 100.00 & 98.56 & ~100.00~ \\ 
Residential & ~90.76~ & ~85.82~ & ~88.81~ & ~98.60~ & ~92.06~ & 78.17 & 83.30 & 94.73 & ~85.45~ \\ 
Commercial & ~79.39~ & ~80.91~ & ~83.29~ & ~82.53~ & ~91.93~ & 80.46 & 88.32 & 85.20 & ~89.36~ \\ 
Road & ~86.21~ & ~79.22~ & ~88.39~ & ~86.87~ & ~92.35~ & 90.93 & 95.37 & 92.90 & ~97.73~ \\ 
Highway & ~62.07~ & ~64.57~ & ~63.13~ & ~61.97~ & ~93.05~ & 55.89 & 63.13 & 85.45 & ~97.88~ \\ 
Railway & ~96.87~ & ~97.81~ & ~92.31~ & ~94.97~ & ~97.34~ & 99.33 & 97.82 & 91.80 & ~97.15~ \\ 
Parking lot 1 & ~93.66~ & ~94.04~ & ~87.22~ & ~90.20~ & ~98.36~ & 99.71 & 97.89 & 91.88 & ~99.52~ \\ 
Parking lot 2 & ~92.28~ & ~90.52~ & ~92.28~ & ~86.67~ & ~90.87~ & 88.77 & 83.51 & 94.87 & ~89.82~ \\ 
Tennis court & ~100.00~ & ~100.00~ & ~99.19~ & ~92.71~ & ~100.00~ & 100.00 & 100.00 & 100.00 & ~100.00~ \\ 
Running track & ~100.00~ & ~100.00~ & ~100.00~ & ~100.00~ & ~100.00~ & 100.00 & 100.00 & 98.75 & ~100.00~ \\\hline
OA & ~88.55~ & ~89.19~ & ~87.70~ & ~88.69~ & ~93.96~ & 89.89 & 90.28 & 92.86 & ~95.11~ \\ 
AA & ~90.76~ & ~90.81~ & ~88.86~ & ~89.94~ & ~94.90~ & 91.39 & 91.68 & 93.76 & ~95.74~ \\ 
Kappa & ~87.59~ & ~88.26~ & ~86.68~ & ~87.75~ & ~93.45~ & 89.02 & 89.44 & 92.10 & ~94.69~ \\ 
\bottomrule\hline
\end{tabular}
\label{houston_compare_table}
\end{table*}

\begin{table*}[ht]
\renewcommand\arraystretch{1.2}
\centering
\caption{Classification performance of different methods on Houston 2018 dataset.}
\begin{tabular}{c|cccccccccc}
\toprule
~~~Class~~~ & FusAtNet & ~TBCNN~ & ~~DFINet~ &  AsyFFNet & ~$S^2$ENet~ & MACN & SS-MAE & MSFMamba & ~DCMNet~\\\hline
Health grass & ~96.28~ & ~94.84~ & ~94.29~ & ~96.90~ & ~90.93~ & 89.53 & 94.00 & ~83.78~ & ~95.79~ \\ 
Stressed grass & ~93.45~ & ~92.60~ & ~92.71~ & ~91.82~ & ~94.76~ & 95.20 & 95.34 & ~94.56~ & ~93.22~ \\ 
Artificial turf & ~100.00~ & ~100.00~ & ~100.00~ & ~100.00~ & ~100.00~ & 100.00 & 100.00 & ~97.97~ & ~100.00~ \\
Evergreen trees & ~98.33~ & ~98.80~ & ~98.92~ & ~99.29~ & ~98.52~ & 99.10 & 98.55 & ~90.48~ & ~99.63~ \\ 
Deciduous trees & ~99.11~ & ~97.12~ & ~97.77~ & ~99.52~ & ~99.39~ & 98.49 & 98.15 & ~79.53~ & ~99.25~ \\ 
Bare earth & ~100.00~ & ~99.61~ & ~100.00~ & ~99.98~ & ~99.96~ & 99.96 & 99.84 & ~98.33~ & ~99.88~ \\ 
Water & ~100.00~ & ~100.00~ & ~100.00~ & ~100.00~ & ~100.00~ & 100.00 & 100.00 & ~95.27~ & ~100.00~ \\ 
Residential buildings & ~97.90~ & ~93.21~ & ~97.36~ & ~95.16~ & ~97.16~ & 95.31 & 94.98 & ~85.76~ & ~97.13~ \\ 
Non-residential buildings & ~93.41~ & ~91.30~ & ~93.29~ & ~95.02~ & ~94.93~ & 93.86 & 96.27 & ~99.47~ & ~95.92~ \\ 
Roads & ~74.56~ & ~61.06~ & ~75.73~ & ~72.72~ & ~74.58~ & 72.11 & 82.97 & ~89.96~ & ~80.92~ \\ 
Sidewalks & ~82.94~ & ~75.91~ & ~83.67~ & ~77.18~ & ~79.47~ & 80.69 & 74.10 & ~79.42~ & ~82.45~ \\ 
Crosswalks & ~84.65~ & ~85.31~ & ~94.23~ & ~93.14~ & ~96.36~ & 97.69 & 89.57 & ~26.85~ & ~98.35~ \\ 
Major thoroughfares & ~86.90~ & ~72.77~ & ~81.20~ & ~84.67~ & ~83.50~ & 79.05 & 83.21 & ~86.80~ & ~87.06~ \\ 
Highways & ~97.59~ & ~95.86~ & ~98.91~ & ~99.53~ & ~98.57~ & 96.86 & 98.08 & ~92.38~ & ~99.85~ \\ 
Railways & ~99.71~ & ~99.78~ & ~99.94~ & ~99.90~ & ~99.84~ & 99.90 & 99.73 & ~98.69~ & ~99.93~ \\ 
Paved parking lots & ~98.08~ & ~90.74~ & ~98.37~ & ~99.18~ & ~97.41~ & 97.74 & 96.36 & ~91.25~ & ~98.76~ \\ 
Unpaved parking lots & ~100.00~ & ~100.00~ & ~100.00~ & ~100.00~ & ~100.00~ & 100.00 & 100.00 & ~98.83~ & ~100.00~ \\ 
Cars & ~97.11~ & ~98.47~ & ~99.09~ & ~96.64~ & ~99.17~ & 98.74 & 97.20 & ~93.75~ & ~98.52~ \\ 
Trains & ~99.63~ & ~99.90~ & ~99.46~ & ~100.00~ & ~100.00~ & 99.74 & 100.00 & ~94.79~ & ~99.91~ \\ 
Stadium seats & ~99.93~ & ~99.92~ & ~99.98~ & ~99.89~ & ~99.98~ & 99.99 & 100.00 & ~99.58~ & ~100.00~ \\\hline
OA & ~91.52~ & ~86.95~ & ~91.03~ & ~91.25~ & ~91.61~ & 90.42 & 92.43 & ~92.38~ & ~93.27~ \\ 
AA & ~94.98~ & ~92.36~ & ~95.25~ & ~95.03~ & ~95.22~ & 94.70 & 94.91 & ~95.51~ & ~96.33~ \\ 
Kappa & ~89.14~ & ~83.39~ & ~88.47~ & ~88.71~ & ~89.17~ & 87.69 & 90.19 & ~90.16~ & ~91.33~ \\ 
\bottomrule
\end{tabular}
\label{houston2018_compare_table}
\end{table*}

\begin{figure*}[htb]
\renewcommand\arraystretch{1.2}
\centering
\includegraphics [width=0.99\textwidth]{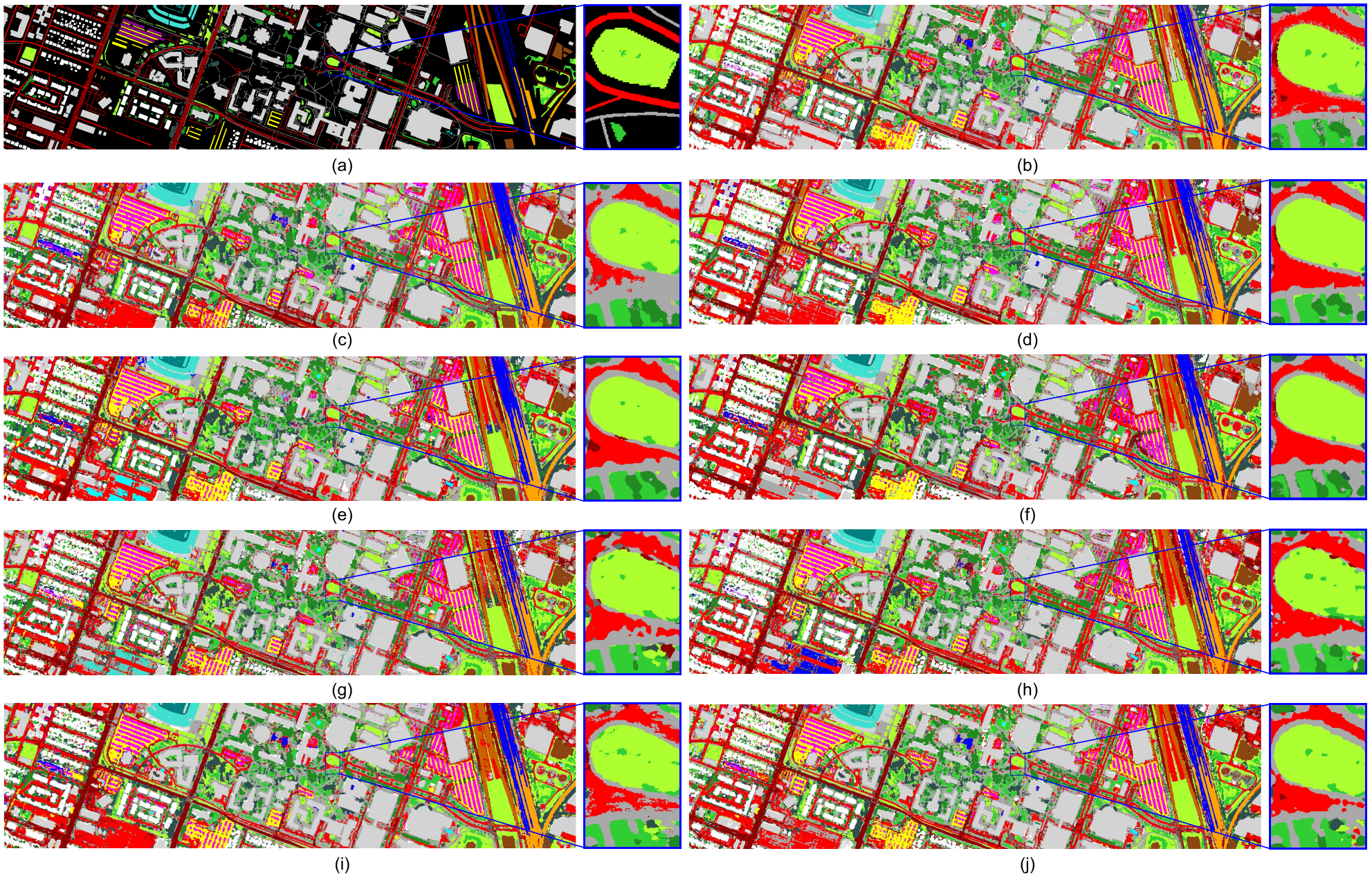}
\caption{Visualized classification maps of different methods on Houston 2018 dataset. (a) Ground Truth. (b) FusAtNet. (c) TBCNN. (d) DFINet. (e) AsyFFNet. (f) $S^2$ENet. (g) MACN. (h) SS-MAE. (i) MSFMamba. (j) DCMNet.}
\label{houston2018_fig_com}
\end{figure*}

\subsection{Parameter Analysis}

\textbf{The Number of Principal Components in HSI.} The number of components $K$ in HSI is an important parameter that affects the classification performance by determining how much spectral information is retained after dimensionality reduction. As shown in Fig. \ref{fig_para_k}, the optimal $K$ values for the Trento, Houston 2018, and Houston 2013 datasets are 10, 20, and 30, respectively. The differences in these values arise from the spectral characteristics of the datasets. For instance, the Trento dataset contains classes with distinct spectral features, allowing a smaller $K$ to retain sufficient information. Conversely, the Houston 2013 dataset has more complex and diverse classes, necessitating a larger $K$ to capture the variations in spectral signatures effectively.

\textbf{The Size of the Input Image Patch.} The size of the input image patch $p$ is a critical parameter for capturing spatial context. A large patch size introduces excessive background noise, while a small patch size may omit critical spatial details. As illustrated in Fig. \ref{fig_para_p}, the optimal patch size varies across datasets: $p = 11$ for the Trento and Houston 2013 datasets, and $p = 13$ for the Houston 2018 dataset. The larger patch size for the Houston 2018 dataset is attributed to its more heterogeneous spatial features, which require broader spatial context for accurate classification.

\textbf{The Feature Sizes in the Routing Space.} The routing space transforms input data into features of size $c \times d$. As shown in Table \ref{table_unify}, the optimal configuration differs across datasets due to their unique spatial and spectral characteristics. For example, in the Trento dataset, the best performance is achieved with $c=32$ and $d=25$, which can be attributed to the dataset's relatively simple and homogeneous nature. In contrast, the Houston 2013 dataset achieves optimal results with $d=9$ and $c=128$, as its complexity benefits from higher channel granularity to enable detailed feature extraction. Meanwhile, the Houston 2018 dataset performs best with $d=25$ and $c=64$, which strikes a balance between spatial resolution and feature depth to accommodate its diverse and complex land cover types. These configurations highlight how the selection of $d$ and $c$ significantly influences the model's ability to extract meaningful features.

\begin{figure*}[ht]
\centering
\includegraphics [width=0.95\textwidth]{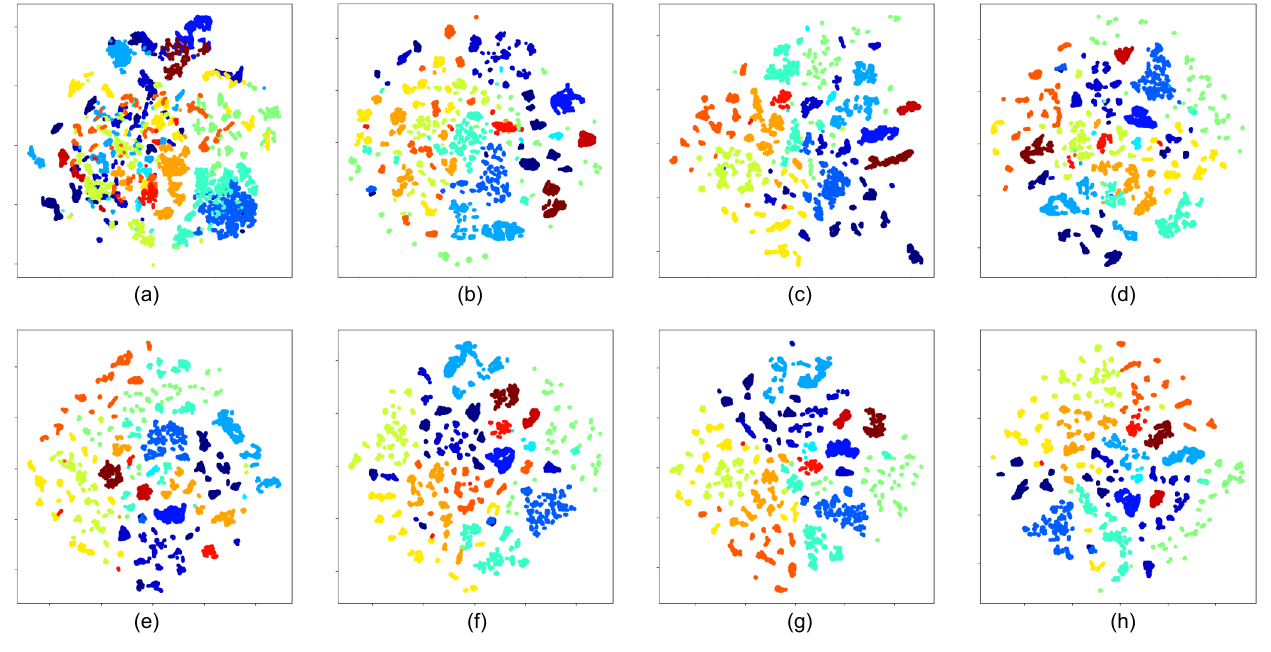}
\caption{Feature visualizations on Houston 2013 dataset by using t-SNE. Different colors indicate different classes. (a) DCMNet-L, (b) DCMNet-H, (c) DCMNet w/o BCAB, (d) DCMNet w/o BSAB  (e) DCMNet w/o bilinear attention, (f) DCMNet with 1-layer routing, (g) DCMNet with 2-layer routing, (h) DCMNet.}
\label{TSNE}
\end{figure*}

\begin{table}[htpb]
\centering
\caption{Performance of The Proposed network with Different Data Sources.}
\renewcommand{\arraystretch}{1.3}
\begin{tabular}{c|ccc}
\hline \toprule
\multirow{2}*{Data} & \multicolumn{3}{c}{Overall Accuracy (\%)} \\ \cmidrule{2-4}
~& ~~ Trento ~~ & Houston 2013 & Houston 2018\\ 
\midrule
~~~DCMNet-L~~~ & 78.02 & 64.89 & 70.17\\ 
~~~DCMNet-H~~~ & 92.75 & 90.61 & 85.88\\ 
~~~DCMNet-HL~~~ & \textbf{98.96} & \textbf{95.11} & \textbf{93.27}\\ 
\bottomrule \hline
\end{tabular}
 \label{table_motality_compare}
\end{table}

\textbf{Evaluation Metrics.} To comprehensively evaluate the performance of the proposed DCMNet and the comparison models, we adopt three widely-used metrics in classification tasks: overall accuracy (OA), average accuracy (AA), and Kappa coefficient. OA represents the percentage of correctly classified pixels across all classes and reflects the overall performance of a model. AA is the mean of the classification accuracies for all classes, highlighting the model's balance in classifying each category. The Kappa coefficient measures the agreement between the predicted classification and the ground truth, considering the possibility of random agreement. These metrics collectively provide a robust assessment of the model's classification ability.

\subsection{Classification Performance}

We compare the classification accuracy of the proposed DCMNet with eight state-of-the-art methods, i.e., FusAtNet \cite{mohla2020fusatnet}, TBCNN \cite{xu2017multisource}, DFINet \cite{gao2022tgrs}, AsyFFNet \cite{li2022asymmetric}, $S^2$ENet \cite{fang2021s2enet}, MACN \cite{li2023mixing}, SS-MAE \cite{lin2023ss}, and MSFMamba \cite{gao2024msfmamba}. In particular, FusAtNet employs an attention mechanism on HSI and Lidar features to highlight spatial and spectral representations, which are then combined to compute modality-specific feature embeddings. TBCNN utilizes a two-tunnel CNN framework to extract spectral-spatial features from HSI and combines them with LiDAR features extracted from a cascade network. DFINet extracts complementary information from multi-source feature pairs using a depth-wise cross-attention module. AsyFFNet adopts weight-shared residual blocks to extract features and uses a feature calibration module for spatial multi-source feature modeling. $S^2$ENet exploits a spatial-spectral enhancement module to realize cross-modal information interaction.  MACN combines self-attention and convolution to efficiently extract spatial and spectral features from multisource remote sensing data. SS-MAE introduces a spatial-spectral masked autoencoder for hyperspectral and LiDAR/SAR data classification. MSFMamba employs Multi-Scale Spatial Mamba (MSpa-Mamba) block, Spectral Mamba (Spe-Mamba) block, and Fusion Mamba (Fus-Mamba) block to achieve multi-source remote sensing classification effectively. For a fair comparison, the training and testing samples for all models are identical. In the data pre-processing stage, we employ data augmentation to improve the robustness of the models and avoid over-fitting. Specifically, we flip image orientation and inject noise randomly into the data to improve the generalization performance.

\textbf{Results on the Trento Dataset.} Table \ref{trento_compare_table} presents the OA, AA, and Kappa of different models on the Trento dataset, with the corresponding classification maps shown in Fig. \ref{trento_fig_com}. The proposed DCMNet achieves the highest overall accuracy (OA), outperforming other models by 0.47\%, 0.35\%, 0.26\%, 0.33\%, 0.95\%, 1.43\%, 0.68\%, and 1.18\% over FusAtNet, TBCNN, DFINet, AsyFFNet, MACN, SS-MAE, and MSFMamba, respectively. In addition to the overall performance, our DCMNet maintains consistently high accuracy across all subclasses, highlighting the strong generalization capability of the dynamic routing mechanism. For instance, FusAtNet, while leveraging attention mechanisms for spatial and spectral feature fusion, tends to overemphasize dominant features, leading to misclassifications in heterogeneous regions such as `Roads'. Similarly, TBCNN, which uses a two-tunnel CNN framework, struggles with fine-grained distinctions in classes like `Apple trees' due to limited cross-modal feature interactions. DFINet performs well in most scenarios but exhibits reduced accuracy in small classes like `Ground' due to its reliance on fixed feature fusion strategies. In comparison, DCMNet dynamically adapts its routing mechanism to prioritize discriminative features across modalities, effectively addressing these challenges. 

\textbf{Results on the Houston 2013 Dataset.} Table \ref{houston_compare_table} shows the overall accuracy (OA), average accuracy (AA), and Kappa score for different models on the Houston 2013 dataset. The classification maps are displayed in Fig. \ref{houston_fig_com}. The results demonstrate that our proposed DCMNet model achieves the best performance in terms of OA and Kappa across the dataset. Specifically, DCMNet improves the OA by 6.56\%, 5.92\%, 7.41\%, 6.42\%, 1.15\%, 5.22\%, 4.83\%, and 2.25\% compared to FusAtNet, TBCNN, DFINet, AsyFFNet, MACN, SS-MAE, and MSFMamba, respectively. Additionally, it is well-documented that 'Roads', 'Highway', and 'Railway' are composed of similar materials but differ in height, making these classes difficult to distinguish using hyperspectral imaging (HSI) alone. However, incorporating LiDAR data significantly aids in this distinction. As indicated in Table \ref{houston_compare_table}, DCMNet achieves an accuracy of over 97\% for these three classes, whereas other models struggle to achieve comparable performance. For example, MSFMamba, with its multi-scale spatial and spectral modeling, performs competitively in `Railway' due to its strong feature extraction capabilities. However, it struggles with `Highway', where feature fusion across modalities is critical for accurate classification. Its fixed fusion strategy limits its adaptability to the varying spatial and spectral characteristics of these classes. Similarly, MACN and SS-MAE, despite their strong performance in spectral feature extraction, fall short in leveraging the structural and height information provided by LiDAR data, resulting in reduced accuracy for `Roads' and `Highway'. The magnified area in Fig. \ref{houston_fig_com} highlights the classification of 'Highway', where DCMNet's results are closest to the ground truth. This improvement can be attributed to the flexibility and robustness of the dynamic routing mechanism utilized in DCMNet.

\textbf{Results on the Houston 2018 Dataset.} Table \ref{houston2018_compare_table} illustrates the OA, AA, and Kappa of different methods on the Houston 2018 dataset. The corresponding classification maps are shown in Fig \ref{houston2018_fig_com}. It can be seen that the proposed DCMNet achieves the best performance compared to other methods. Specifically, our proposed DCMNet improves OA by 1.75\%, 6.32\%, 2.24\%, 2.02\%, 1.66\%, 2.85\%, 0.84\%, 0.89\% over FusAtNet, TBCNN, DFINet, AsyFFNet, MACN, SS-MAE, and MSFMamba, respectively. The Houston 2018 dataset includes highly heterogeneous classes such as `Health grass' and `Stressed grass', which require both spatial and spectral discrimination. MSFMamba benefits from its multi-scale modules, but its feature fusion strategy lacks the adaptability to handle such diversity effectively. DCMNet’s dynamic routing mechanism excels in these scenarios by dynamically balancing spatial and spectral information, resulting in finer distinctions between similar classes.

\textbf{Analysis of Suboptimal Performance in Specific Classes.}
Although DCMNet achieves the best overall performance across all three datasets, certain classes exhibit slightly suboptimal results due to inherent challenges in data distribution and class characteristics. For example, in the Houston 2013 dataset, the class `Residential' shows slightly lower accuracy compared to other classes. This can be attributed to the high intra-class variability within `Residential', such as diverse building materials and roof shapes, which introduce spectral inconsistencies. Despite this slight reduction in performance, it is important to emphasize that DCMNet's accuracy for `Residential' remains highly competitive and does not significantly lag behind other methods. Furthermore, DCMNet compensates for this by achieving superior overall performance across all crosses, demonstrating its robustness and adaptability in handling diverse scenarios.

\subsection{Ablation Study}

We conduct several ablation studies to evaluate the effectiveness of the DCMNet on the task of HSI and LiDAR data classification. Specifically, we first reveal the impact of data from different modalities on the classification performance, i.e., only the LiDAR modality, only the HSI modality, and both modalities. Then, we investigate the impact of different components of the network structure on the classification performance, i.e., different blocks and bilinear attention. Finally, we examine how the number of layers affects the classification performance. 

\textbf{Analysis of Different Modalities.} To explore the effect of the complementary information of the two modalities on the classification performance, we show the overall accuracy of the models trained by LiDAR data, HSI, and data from both modalities in Table \ref{table_motality_compare}. It can be observed that the overall accuracy of the model trained with LiDAR data (DCMNet-L) is extremely low because it is unable to distinguish objects of similar heights. Although the model trained with HSI data (DCMNet-H) can get a tolerable overall accuracy, it cannot effectively distinguish between two different classes composed of similar materials. Model trained by both HSI and LiDAR data (DCMNet-HL) achieves the best performance. It is evident that the complementary information between HSI and LiDAR data is fully exploited.

\textbf{Effects of Feature Interactive Blocks.} In Table \ref{table_cell_compare}, we evaluate the effect of three feature interactive blocks on the classification performance. For Trento dataset, BSAB plays a more significant role in classification than BCAB, while for Houston 2013 and Houston 2018 datasets, BCAB plays a more significant role in classification performance than BSAB. This phenomenon may be caused by the difference in the characteristics of the ground objects in these datasets. Taking the `Woods' class in Trento dataset as an example, its height and texture information are somewhat similar to other classes. However, its LiDAR data shows the unevenness in its spatial distribution, while BSAB would be more suitable for such distribution. In addition, when ICB is missing, the accuracy improvement is limited or even decreases. The reason is that complicated feature interactions may not be essential for simple images. When three blocks are employed, our DCMNet achieves the best classification performance, which proves the necessity of three feature interactive blocks. 

\textbf{Effects of Dynamic Routing Mechanism.}
To further explore the impact of the dynamic routing mechanism, we compare two simplified versions of DCMNet in Table \ref{table_dr_compare}. DCMNet$'$ removes the router and directly averages the features from all blocks. DCMNet*, on the other hand, simplifies the model even further by removing both the router and the feature interactive blocks. The performance differences between these versions demonstrate the effectiveness of the dynamic routing mechanism. DCMNet$'$ exhibits a noticeable performance drop compared to the full DCMNet, especially on the Houston 2018 dataset, where the accuracy decreases from 93.27\% to 92.19\%. This suggests that the router plays a critical role in dynamically integrating feature information, improving classification performance.  However, on the Trento dataset, the performance drop for DCMNet$'$ is less pronounced (from 98.96\% to 98.92\%), likely because the Trento dataset is relatively simple. It does not require specialized handling of spatial or spectral features to achieve good classification results. DCMNet*, which further simplifies the architecture, results in a larger performance drop across all datasets, with the Houston 2018 accuracy falling to 89.69\%. These results confirm that the dynamic routing mechanism is essential for optimizing the flow of information between the blocks and maintaining high classification performance.

\textbf{Analysis of Computational Complexity.}
In terms of parameters, GFLOPs, and computational times on Houston 2018 dataset (see Table \ref{complexity_compare}), DCMNet maintains a balance between complexity and performance. Despite including many components, DCMNet demonstrates reasonable computational demands with training and inference times of 0.0274s and 0.0097s, respectively, comparable to simpler models like $S^2$ENet. The simplified versions, DCMNet$'$ and DCMNet*, show reductions in computational complexity by removing the router and the feature interactive blocks. However, as shown in the performance comparison, this reduction comes with a noticeable loss in accuracy, particularly in DCMNet*. These findings indicate that while dynamic routing introduces some complexity, it has a minimal impact on computational cost and is justified by the performance improvements it provides.

\begin{table}[h]
\centering
\caption{Influence of BCAB, BSAB, and ICB on classification accuracy of the proposed DCMNet.}
\renewcommand{\arraystretch}{1.3}
\begin{tabular}{ccc|ccc} \hline
\toprule        
\multirow{2}*{BCAB} & \multirow{2}*{BSAB} & \multirow{2}*{~ICB~} & \multicolumn{3}{c}{Overall Accuracy (\%)} \\
\cmidrule{4-6}
 &  &  & ~ Trento ~ & Houston 2013 & Houston 2018 \\ 
\midrule
\checked & $\times$ & $\times$ & 96.67 & 92.91 & 88.56\\ 
$\times$ & \checked & $\times$ & 98.39 & 88.76 & 90.11 \\ 
$\times$ & $\times$ & \checked & 98.58 & 93.34 & 91.25\\ 
\checked & \checked & $\times$ & 96.91 & 90.75 & 91.37\\ 
\checked & $\times$ & \checked & 98.67 & 94.15 & 92.74\\ 
$\times$ & \checked & \checked & 98.81 & 93.61 & 92.41\\ 
\checked & \checked & \checked & ~\textbf{98.96}~ & ~\textbf{95.11}~ & \textbf{93.27}~\\ 
\bottomrule \hline
\end{tabular}
\label{table_cell_compare}
\end{table}

\begin{table}[htb]
\centering
\caption{Influence of dynamic routing mechanism on classification performance of the proposed DCMNet.}
\renewcommand{\arraystretch}{1.3}
\begin{tabular}{c|ccc} 
\hline\toprule       
\multirow{2}*{Method} & \multicolumn{3}{c}{Overall Accuracy (\%)} \\
\cmidrule{2-4}
~ & ~~ Trento ~~ & Houston 2013 & Houston 2018 \\
\midrule
DCMNet & 98.96 & 95.11 & 93.27 \\
DCMNet$'$ & 98.92 & 94.20 & 92.19 \\
DCMNet* & 97.49 & 92.78 & 89.69 \\ 
\bottomrule\hline
\end{tabular}
\label{table_dr_compare}
\end{table}


\begin{table}[t]
\renewcommand\arraystretch{1.4}
\centering
\caption{Quantitative Comparison of Model Complexity in Terms of the Number of Parameters and GFLOPs on Houston 2018 Dataset.}
\begin{tabular}{c|ccc}
\hline\toprule
Model & ~Params. (MB)~ & GFLOPs & \makecell{Training/Inference time\\per sample} \\\hline
FusAtNet & 36.9854 & 4.7384 & 0.4843 / 0.2147 \\ 
TBCNN & 0.3875 & 0.0063 & 0.0098 / 0.0038 \\ 
DFINet & 3.0394 & 0.0859 & 0.0237 / 0.0091 \\ 
AsyFFNet & 4.0558 & 0.1352 & 0.0285 / 0.0149 \\
$S^2$ENet & 0.4892 & 0.0216 & 0.0116 / 0.0040 \\
SS-MAE & 4.2352 & 0.1423 & 0.0304 / 0.0111 \\
MACN & 2.2399 & 0.2472 & 0.0231 / 0.0084 \\ 
MSFMamba & 1.4591 & 0.0371 & 0.0152 / 0.0058 \\
DCMNet & 3.8262 & 0.0459 & 0.0274 / 0.0097 \\ 
DCMNet$'$ & 3.6710 & 0.0426 & 0.0269 / 0.0081 \\
DCMNet* & 2.1363 & 0.0306 & 0.0194 / 0.0068 \\
\bottomrule\hline
\end{tabular}
\label{complexity_compare}
\end{table}

\textbf{Analysis of Bilinear Attention.} To investigate the influence of second-order feature interactions in BCAB and BSAB, we compare the classification results of the model with self-attention and bilinear attention. As shown in Table \ref{table_bp_compare}. It is noticed that the accuracy of the latter is 0.34\%, 0.50\%, and 0.29\% higher than that of self-attention on Trento, Houston 2013, and Houston 2018 datasets, respectively.

\begin{table}[htb]
\centering
\caption{Influence of bilinear attention on classification performance of the proposed DCMNet.}
\renewcommand{\arraystretch}{1.3}
\begin{tabular}{c|ccc} 
\hline\toprule       
\multirow{2}*{Method} & \multicolumn{3}{c}{Overall Accuracy(\%)} \\
\cmidrule{2-4}
~ & ~~ Trento ~~ & Houston 2013 & Houston 2018 \\
\midrule
Self-attention & 98.52 & 94.61 & 92.98\\
Bilinear attention & \textbf{98.96} & \textbf{95.11} & \textbf{93.27}\\ 
\bottomrule\hline
\end{tabular}
 \label{table_bp_compare}
\end{table}

\begin{table}[htb]
\centering
\caption{Influence of layer numbers in dynamic routing on classification performance of DCMNet.}
\renewcommand{\arraystretch}{1.3}
\begin{tabular}{c|ccc} 
\hline\toprule     
\multirow{2}*{Number of layers} & \multicolumn{3}{c}{Overall Accuracy (\%)} \\
\cmidrule{2-4}
~ & Trento & Houston 2013 & Houston 2018 \\
\midrule
~~~1~~~ & 96.39 & 93.18 & 91.33\\ 
~~~2~~~ & 98.51 & 94.65 & 92.82\\ 
~~~3~~~ & \textbf{98.96} & \textbf{95.11} & \textbf{93.27}\\ 
\bottomrule\hline
\end{tabular}
 \label{table_ln_comp}
\end{table}

\textbf{Number of Layers in Dynamic Routing.} In order to evaluate the network depth of the proposed model, we study the number of layers in the proposed DCMNet. The experimental results are shown in Table \ref{table_ln_comp}. It can be observed that the DCMNet with 3 layers achieves the best classification results. The reason is that more feasible feature combinations are available, when the number of layers increases. Therefore, the cross-modal feature interactions and representations are enhanced.

\begin{figure}[htb]
\centering
\includegraphics [width=2in]{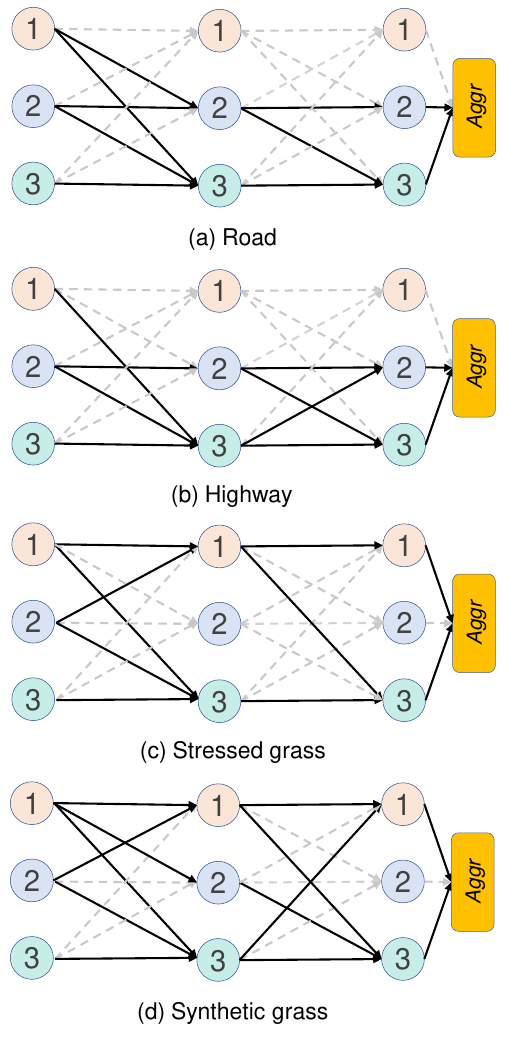}
\caption{Routing path visualization of several examples.}
\label{routing_vis}
\end{figure}

\textbf{Feature Visualization.} To more intuitively evaluate the performance of DCMNets with different combinations of network components, we map the features obtained on Houston 2013 dataset into the two-dimensional space, and the results are shown in Fig. \ref{TSNE}. DCMNet-L denotes the model trained by LiDAR. DCMNet-H denotes the model trained by HSI.  As can be observed, the features of the proposed DCMNet with all components display the most compact, well-defined clusters. Hence, the feature visualizations also demonstrate the effectiveness of each component in our DCMNet.

\textbf{Routing Visualization.} To intuitively explore the effect of the dynamic routing in our DCMNet, we visualize the routing path of several examples. The path above a threshold of 0.3 is denoted as solid line. We select `Stressed grass', `Synthetic grass', `Road', and `Highway' from Houston 2013 dataset. `Stressed grass' and `Synthetic grass' are two classes with similar heights and different compositions, BSAB and ICB are more involved in the feature extraction. The reason may be that textures play a crucial role in distinguishing between `Stressed grass' and `Synthetic grass'. Additionally, For `Road' and `Highway' classification, BCAB and ICB are more critical. It further demonstrates that spectral information is more important for `Road' and `Highway' classification. Finally, it is noteworthy that ICB is the most frequently used in all paths. It is evident that for the Houston 2013 dataset, basic feature interactive is essential.

\subsection{Limitations of the Proposed DCMNet}

While the proposed DCMNet demonstrates strong performance in multi-source remote sensing classification tasks, several limitations and challenges remain that need further investigation.

One potential limitation is the model's capability to handle multi-source data from sensors with different resolutions. DCMNet relies on a dynamic routing mechanism to integrate heterogeneous data, but significant differences in spatial or spectral resolution between sensors may pose challenges in feature alignment and fusion. Specifically, for features from different sources that are projected into the routing space, large differences in spatial resolution may make it difficult for a simple projector to effectively align them within a common space. This could lead to performance degradation. Our future work could explore resolution-adaptive techniques to better handle cross-modal data with varying resolutions, ensuring more robust feature fusion across different sensors.

Another limitation involves extending DCMNet to other types of remote sensing data, such as radar or thermal imagery. While DCMNet has been validated on HSI and LiDAR datasets, applying this framework to other modalities may require adjustments to accommodate the unique characteristics of these data types, such as the non-visual nature of radar or the infrared spectrum in thermal imagery. Future work could focus on adapting DCMNet to different sensor modalities by incorporating modality-specific preprocessing techniques or developing more generalized feature extraction modules to ensure its effectiveness across diverse data sources.

\section{Conclusion}

In this paper, we propose DCMNet to achieve an adaptive feature interaction path based on the input data for HSI and LiDAR joint classification. Concretely,  in order to effectively capture the complementary cross-modal features, we construct a three-layer routing space in the fully-connected manner. Furthermore, we design three feature interactive blocks to compute spatial, spectral, and discriminative features, respectively. Bilinear attention is introduced to enhance feature interactions for spatial and channel feature representation. Finally, routing gate is designed in each block, and each block can selectively pass the signals to the blocks in the next layer. Such dynamic routing mechanism is employed to find the best paths for different input data. Extensive experimental results on Trento dataset, Houston 2013 dataset, and Houston 2018 datasets demonstrate the effectiveness and superiority of our DCMNet. We also visualize the feature patterns and representative routing paths, which also show that our DCMNet can generate better feature representations.

\bibliography{source}
\bibliographystyle{IEEEtran}

\end{document}